\documentclass[aps,prd,preprint,showpacs,floatfix,preprintnumbers,nofootinbib,superscriptaddress,showkeys,longbibliography
]{revtex4}
\usepackage[utf8]{inputenc}
\usepackage{bm}
\usepackage{slashed}
\usepackage{fancybox}
\usepackage{hhline}
\usepackage{dcolumn}
\usepackage{textcomp}
\usepackage[toc,page]{appendix}
\usepackage{epsfig,graphics,graphicx}
\usepackage{amsfonts,amssymb,amsmath}
\usepackage{tikz}
\usepackage{tikz-feynman}
\usetikzlibrary{shapes,arrows,positioning,decorations,backgrounds,calc,er,patterns}
\newcommand{\btik}{\begin{tikzpicture}}
\newcommand{\etik}{\end{tikzpicture}}
\usepackage{pifont}
\usepackage{bm}
\usepackage{appendix}
\usepackage[mathscr]{euscript}
\usepackage{mathrsfs}
\usepackage{multirow}
\usepackage{rotating}
\usepackage{adjustbox}
\usepackage{caption}
\usepackage{booktabs}
\usepackage{color}   

\newcommand{\beq}{\begin{equation}}
\newcommand{\eeq}{\end{equation}}
\newcommand{\bea}{\begin{eqnarray}}
\newcommand{\eea}{\end{eqnarray}}

\newcommand{\eps}{\epsilon}
\newcommand{\ord}[1]{{\cal{O}}( #1 )}
\newcommand{\proj}[1]{{\cal{P}}_{#1}}
\newcommand{\B}{{\bf B}}
\newcommand{\Bdag}{{\bf B^\dagger}}

\newcommand{\gA}{\mathring g_A}
\newcommand{\gAoF}{\frac{\gA}{F_\pi}}

\newcommand{\lepspi}{\frac{\lambda_\eps}{(4\pi)^2}}

\DeclareFontFamily{OT1}{pzc}{}
\DeclareFontShape{OT1}{pzc}{m}{it} 
{<-> s * [0.900] pzcmi7t}{}
\DeclareMathAlphabet{\mathpzc}{OT1}{pzc} 
{m}{it}
\DeclareMathAlphabet{\mathcalligra}{T1}{calligra}{m}{n}
\usepackage{natbib,hyperref}
\usepackage{url}
\hypersetup{pdfauthor={me}, colorlinks=true, citecolor=blue, urlcolor=blue, linkcolor=black}
\begin{document}


\title{Pion-Nucleon Scattering in Baryon Chiral Perturbation Theory combined with the \texorpdfstring{$\bm{ 1/N_c}$ }{1/Nc}Expansion}

\author{D.~Jayakodige}
\email[ E-mail: ]{dulitha@jlab.org}
\affiliation{Department of Physics, Hampton University, Hampton, VA 23668, USA}

\author{J.~L.~Goity}
\email[ E-mail: ]{goity@jlab.org}
\affiliation{Department of Physics, Hampton University, Hampton, VA 23668, USA}
\affiliation{Theory Center, Jefferson Lab, Newport News, VA 23606, USA}

\begin{abstract}
 This work implements the combined BChPT and $1/N_c$ expansions for pion-nucleon elastic scattering. The effective theory is based on the baryon sector dynamical spin-flavor $SU(4)$ symmetry emergent in the large $N_c$ limit, whose breaking is controlled by the $1/N_c$ expansion. The non-commutativity of the chiral and $1/N_c$ expansions in unitarity corrections (loops) requires a linking of both expansions. As it was shown in the case of baryon masses and currents, the natural linking is the $\xi$-expansion, in which $\ord{p}=\ord{1/N_c}=\ord{\xi}$. The spin-flavor symmetry requires that the ground state baryons span an $SU(4)$ symmetric irreducible representation which implies that in particular $N$ and $\Delta$  are active degrees of freedom in the effective theory. 
 The scattering amplitude is expanded to the next-to-next-to leading order in the $\xi$ expansion, corresponding to the  one-loop contributions with the LO Lagrangian. The results are given for generic $N_c$ in order to demonstrate the consistency of the framework. The spin-flavor symmetry plays a central role in maintaining the consistency of the effective theory with respect to the $1/N_c$ expansion. This consistency manifests itself in an improvement in the convergence of the low energy expansion with respect to the case of the ordinary BChPT without an explicit dynamical $\Delta$, which is known to be  inconsistent with the constraints of $N_c$ scaling.   
 Fits to the $\pi N\to \pi N$  S, P and D   partial wave amplitudes from the SAID data base  are finally used to test the framework and to determine the energy range of its applicability. 
\end{abstract}


\preprint{JLAB-THY-25-4343}

\maketitle

\section{Introduction} \label{sec:introduction}

The formulation of baryon effective theories has spanned different frameworks, starting with the manifestly relativistic baryon chiral perturbation theory (BChPT)
\cite{Gasser:1987rb,Krause:1990xc}, followed by the same relativistic framework endowed with the infrared regularization scheme \cite{Ellis:1997kc,Becher:1999he,Schindler:2003xv}. Those formulations were implemented with only the spin-1/2 baryons, and later extended to include the spin-3/2 baryons \cite{Hacker:2005fh}. At the same time, a formulation  was developed based on the expansion in $1/m$, $m$ being the baryon mass \cite{Jenkins:1990jv,Jenkins:1991es,Jenkins:1991ne}, known as the heavy BChPT or HBChPT. Within this framework,  and   including  the spin 3/2 baryons, it was observed that the convergence of the low energy expansion as shown by the  one-loop corrections, is greatly improved due to partial cancellations between the contributions of spin-1/2 and spin-3/2 baryons in loops \cite{Jenkins:1991ne,Jenkins:1991es}. The poor convergence of the low energy expansion with only spin-1/2 baryons is primarily due to the large $\pi N$ coupling $g_{\pi NN}\simeq 13.47$. Those observed improvements are  result of a fundamental aspect of QCD, namely the fact that, at sufficiently large $N_c$, QCD admits an expansion in powers of $1/N_c$ \cite{'tHooft:1973jz}. In the large $N_c$ limit, $g_{\pi NN}=\ord{N_c^{3/2}}$, while observables such as the $\pi N$ scattering amplitudes  must remain of zeroth order in $N_c$. This leads to   dynamical constraints that must be satisfied at large $N_c$ \cite{Gervais:1983wq,Dashen:1993ac}. Those constraints can be associated with an emergent (contracted) spin-flavor symmetry $SU(2N_f)$  in the baryon sector, $N_f$ being the number of light flavors, which in particular requires the inclusion of higher spin baryons, namely the $\Delta$ in the physical case $N_c=3$. In such a framework, the $1/N_c$ expansion can be systematically implemented, providing in particular the deviations from the spin-flavor symmetry as sub-leading effects in $1/N_c$. The spin-flavor symmetry is already explicit in the Skyrme model \cite{Adkins:1983ya}, and to no surprise it also emerges in the large $N_c$  quark model for baryons, justifying the $SU(4)$ and $SU(6)$ symmetries introduced in the 1960's. Note that the spin-flavor symmetry cannot be realized relativistically, as demonstrated by the Coleman-Mandula theorem \cite{coleman1967all}. Since at large $N_c$  baryons have masses $\ord{N_c}$,   and the interest is in studying the low energy limit,  where the dynamically relevant momenta and energies are small, the natural reference frame to be used is that in which the baryon has a small 3-momentum, and thus it is almost at rest. Thus, a non-relativistic framework for baryons is the natural one, where the spin-flavor symmetry can be indeed realized.

	Endowing BChPT with the consistency requirements of the large $N_c$ limit is therefore a natural approach. Following on the initial work \cite{Jenkins:1995gc}, the implementation of the effective theory at the one-loop level with the corresponding renormalization   was initiated \cite{CalleCordon:2012xz}, where baryon masses and axial currents were studied, followed by studies for three flavors \cite{Fernando:2017yqd}. In the latter references the combined chiral and $1/N_c$ expansions were introduced, in which both expansions are linked according to $\ord{p}=\ord{1/N_c}$, a scheme  coined as the $\xi$-expansion. Such linking is necessary in practice because the different observables contain non-analytic terms involving their ratios. In particular, for such terms the $1/N_c$ expansion is very slowly convergent as it becomes weighted with respect to the small chiral scales, a feature known as the non-commutativity of the expansions. This framework based on the $\xi$-expansion, hereafter referred to as $\rm{BChPT\times 1/N_c}$, has been applied to masses and currents \cite{CalleCordon:2012xz,Fernando:2017yqd}. Further testing is here carried out for the case of $\pi-$baryon scattering. As the most studied process in the different versions of effective theory \cite{Gasser:1987rb,Mojzis:1997tu,Fettes:1998ud,Fettes:2000xg,Fettes:2000bb,Fettes:2001cr,Becher:2001hv,Torikoshi:2002bt,Alarcon:2011zs,Alarcon:2012kn,Chen:2012nx,Hoferichter:2015hva,Yao:2016vbz}, and empirically as the most accurately known, it is   ideal for that purpose. 
	
	The first objective of this work is to implement   $\rm{BChPT\times 1/N_c}$ at the one-loop level, i.e.  next-to-next-leading-order (NNLO) in the $\xi$ expansion,   for the  $\pi B\to \pi B'$ scattering amplitude in the isospin symmetry limit, carrying out the renormalization at generic $N_c$. The second objective is to confront the results with the $\pi N\to \pi N$ amplitudes as provided by the SAID data base \cite{Arndt:2006bf,SAID-webpage}.

	The work is organized as follows:   Section \ref{sec:BChPTx1/Nc} presents the ${\rm BChPT\times 1/N_c}$  framework,
	Section \ref{sec:masses-axial-couplings} presents the NNLO calculation of the masses and $\pi$-baryon couplings,
Section \ref{sec:scattering-amplitudes} presents the calculation of the $\pi$-baryon amplitudes to the NNLO,
Section \ref{sec:piN-piN-fits} presents the fits to the $\pi N\to \pi N$ S, P, and D partial wave scattering amplitudes from the SAID data base, and 
Section \ref{sec:conclusions}  presents   a summary and conclusions.
	Finally, the appendices contain most of the computational details and explicit results, offering readers adequate information to reproduce the results.

\section{$\boldsymbol{\rm BChPT  \times  1/N_c }$}\label{sec:BChPTx1/Nc}

In the large $N_c$ limit, QCD must admit an expansion in powers of $1/N_c$. This is evident at the level of QCD Feynman diagrams as 'tHooft showed 50 years ago \cite{'tHooft:1973jz}. It is expected that the expansion holds at the non-perturbative level, and thus it ought to be implemented at the hadronic level, in particular in effective theories.
 
 The $1/N_c$ expansion requires a proper definition. The one that is most realistic for purposes of phenomenology is 'tHooft's expansion \cite{'tHooft:1973jz}, where the number of flavors $N_f$ is kept fixed and quarks are in the fundamental irreducible representation (irrep) of $SU(N_c)$, with the usual Standard Model isospin and the assignments of Hypercharges such that, for arbitrary $N_c$, the SM quantum numbers of the mesons and baryons identified with the physical ones are left unchanged. Since the expansion actually compares different theories, i.e. with different numbers of degrees of freedom, the defining scales of QCD must be prescribed for each $N_c$. It is convenient to do so with hadronic scales, namely the masses of the ground state mesons, $\pi$, $K$, which are most sensitive to quark masses, and $m_\rho$ for setting the corresponding value of the QCD scale, which along with the quark masses will present sub-leading $1/N_c$ dependency. 
 
 Meson masses scale as $\ord{N_c^0}$, while baryon masses are $\ord{N_c}$, and the meson decay constants, in particular $F_\pi$, are quantities $\ord{\sqrt{N_c}}$. The meson-meson interactions are suppressed in the large $N_c$ limit  with amplitudes scaling as $\ord{1/N_c^{n/2}}$, where $n$ is the number of initial plus final mesons in the interaction. On the other hand,  the meson-baryon couplings can even grow with $N_c$, as it is the case of the $\pi$-baryon coupling $g_{\pi BB}$, which is proportional to $N_c^{3/2}$ \cite{Witten:1979kh}. That scaling of $g_{\pi BB}$ and the   requirement of a  finite $\pi$-baryon scattering amplitude in the large $N_c$ limit as required by  unitarity, implies the emergence of a dynamical contracted $SU(2N_f)$ spin-flavor symmetry for baryons \cite{Gervais:1983wq, Dashen:1993as,Dashen:1993ac}. Baryon states must then form multiplets of that symmetry. In particular, for $N_f=2$, the ground state baryons are in the totally symmetric $SU(4)$ multiplet with $S=I=1/2,\cdots,N_c/2$ ($N_c$ odd). The states in the multiplet with $S=\ord{N_c^0}$ must have mass splitting $\ord{1/N_c}$, as required by the aforementioned consistency. The hypothesis that the $1/N_c$ expansion holds down to $N_c=3$ implies that the nucleon and $\Delta$ resonance must belong to the ground state spin-flavor multiplet, and must be active degrees of freedom in the low energy effective theory. Appendix \ref{app:su4} gives   details on $SU(4)$.
 
 The  low energy effective theory must be consistent with chiral symmetry and the $1/N_c$ expansion. In the pure Goldstone Boson sector this was formulated long ago \cite{Schechter:1979bn,DiVecchia:1980yfw,Witten:1980sp,Herrera-Siklody:1996tqr,Kaiser:2000gs}, and later it was extended to baryons \cite{Jenkins:1995gc, FloresMendieta:2006ei, CalleCordon:2012xz}. For baryons, the implementation is along the following  lines: i) The active fields in the Lagrangian are the Goldstone Bosons and the  baryon ground-state spin-flavor multiplet. ii) The baryon masses being $\ord{N_c}$, the $1/N_c$ expansion requires the use of the non-relativistic baryon fields. iii) The effective Lagrangians must be manifestly invariant under chiral $SU_L(2)\times SU_R(2)$  as well as under the pertinent continuous and  discrete space-time symmetries. iv) The baryon Lagrangians consist of a composition of a chiral tensor and a spin-flavor tensor built with products of the spin-flavor generators. v) The spin-flavor tensors must be such that the constraints of the large $N_c$ limit are not violated in observables. The necessary details are provided in the Appendices, where Appendix \ref{app:ChBBs} gives  the chiral building blocks, Appendix \ref{app:su4} gives a summary of the $SU(4)$ Algebra and matrix elements in the symmetric $SU(4)$ irrep, and Appendix \ref{app:SFOperators} gives the bases of spin-flavor tensors of definite spin and isospin, as needed in this work.
 
 The   structure of the baryon Lagrangians is therefore of the following general form:
 	\beq
 {\cal{L}}\sim \Bdag T_{\chi} T_{SF} B,
 \eeq
 where $\B$ is the baryon spin-flavor multiplet, $T_{\chi}$ is the chiral tensor formed with the chiral building blocks, and $T_{SF}$ is the spin-flavor tensor which is a composite operator of products of  $SU(4)$ generators. Each Lagrangian term is chirally invariant, their chiral order is determined by $T_{\chi}$, while the leading in $1/N_c$ order is determined by $T_{SF}$. The Lagrangian will contain   leading and sub-leading orders in $1/N_c$, the latter from the expansion of the chiral tensor through the factors of $1/F_\pi$ multiplying the pion fields. The spin-flavor tensors are defined to contain the $1/N_c$ suppression factor determined by the n-bodyness of the tensor, in this way the LEC in front of each Lagrangian term is $\ord{N_c^0}$, followed by sub-leading corrections. This in particular implies the obvious fact  that fully determining the effective theory for arbitrary $N_c$ requires explicit  knowledge of QCD at different $N_c$, which  could be achieved using lattice QCD. This is however no significant hindrance to the applications to the real world with $N_c=3$.
 
The low energy effective theory contains two small scales, namely the small energy/momenta characteristic of the chiral expansion and the $\ord{1/N_c}$ baryon mass splittings, e.g. $\Delta=m_\Delta-m_N$. Observables contain dependencies on ratios of both scales, e.g. $M_\pi/\Delta$, which preclude the independent low energy and $1/N_c$ expansions. Such dependencies appear in the loop contributions involving GB and baryon propagators. This "non-commutativity" of the expansions demands that either only one of them is implemented, or they be linked. The latter is evidently the option that works best for the real world as it was shown in several works \cite{CalleCordon:2012xz, Fernando:2017yqd}, in which the $\xi$-power counting scheme was introduced, according to which $\ord{p}=\ord{1/N_c}=\ord{\xi}$. Indeed, non-analytic terms  stemming from loop corrections or pole terms in amplitudes   that involve ratios of the form $p/(1/N_c)$ are slowly convergent in either expansion, and need to be kept at face value. Lagrangians are, therefore, organized by their order in the $\xi$ expansion, and as shown later, the UV divergencies of loop contributions to observables are analytic in $1/N_c$ and in particular in powers of $\xi$.


While the baryon Lagrangians needed for this work are given in Appendix \ref{app:Lagrangians}, it is convenient to present here the LO one, that provides the interactions for the one-loop calculations:
\begin{widetext}
\beq
\mathcal{L}_B^{(1)}=\Bdag\Big(i D_0+\gA u^{i a} G^{i a}-\frac{C_{H F}}{N_c} \hat S^2+\frac{c_1 N_c}{2 \Lambda} \langle{\chi}_{+}\rangle\Big) \B, 
\eeq
\end{widetext}
which is expressed in terms of the chiral building blocks in Appendix \ref{app:ChBBs}. The first term contains the residual energy of the baryon in the heavy mass expansion and   two-pion vertices (Weinberg-Tomozawa terms). The second term gives the single pion coupling to the baryons in addition to the axial current coupling. The third term gives the residual mass contribution to the baryon according to its spin, giving mass splittings $\ord{1/N_c}$ between baryons with spin $\ord{N_c^0}$. The final term gives the LO quark mass contribution to the baryon mass, and also two-pion vertices.  The arbitrary scale $\Lambda$ is introduced for convenience to have $c_1$ dimensionless, as well as in the higher order Lagrangians to have dimensionless LECs, and will be set equal to $m_\rho$ in the explicit calculations. The axial coupling $\gA$ is at LO related  to the axial coupling $g_A$ of the nucleon at $N_c=3$ by $\frac 56 \gA=g_A=1.267$. For notational convenience, the mass shift operator is defined by $\delta\hat m=\frac{C_{H F}}{N_c} \hat S^2$, where the spin-flavor singlet contribution from the quark masses is omitted, since only baryon mass differences appear in loop diagrams.
Appendix \ref{app:Lagrangians} Table \ref{vertices} gives the vertices from $\mathcal{L}_B^{(1)}$ needed in this work.

The NLO and NNLO calculations involve the one-loop diagrams. Their power counting is defined by the resulting terms analytic in chiral and $1/N_c$ powers, the $\xi$ power   being the sum of them. In general, individual diagrams containing vertices proportional to $\gA$ will give contributions that violate  large $N_c$ consistency. Such terms must then be canceled when adding all the pertinent diagrams. These cancellations  provide in particular  for   useful checks  of the  calculations.

\section{Baryon Masses and axial couplings}\label{sec:masses-axial-couplings}

\subsection{Masses}
At LO the baryon mass formula is simply:
\beq
m_B(S)=m_0+\frac{C_{\rm HF}}{N_c} S(S+1)-{2} c_1 N_c \frac{M_\pi^2}{\Lambda},
\eeq
where $m_0$ is the common mass $\ord{N_c}$, $C_{\rm HF}$ is determined from the $\Delta-N$ mass difference, and $c_1$ gives the dependence on $M_\pi$, which is proportional to $N_c$. Indeed, the $\sigma$ term is in effect a quantity $\ord{N_c}$, which implies that even at large $N_c$ a finite fraction of the baryon mass is due to the current quark masses.

At NNLO the masses include the one-loop terms shown in Fig.(\ref{one-loop-mass}) and counterterms (CT). 
\begin{widetext}
\begin{center}
\begin{figure}[hbt!]
\centerline{\includegraphics[width=14.cm,angle=-0]{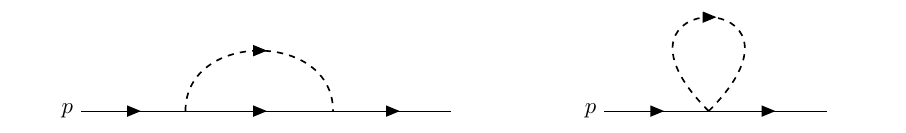}}
\caption{One-loop contributions to the baryon self-energy}
\label{one-loop-mass}
\end{figure}
\end{center}
The one-loop contribution to the self energy operator reads:
\bea
\hat\Sigma(p^0)&=& i \Big(\left( \frac{\gA}{F_\pi}\right)^2 \sum_n G^{ia}\proj{n}G^{ia}\; {\cal{I}}(\delta m_n-p^0,M_\pi) 
  +2    N_c c_1 \frac{M_\pi^2}{\Lambda F_\pi^2} \cal{I}(M_\pi)\Big) ,
\eea
where $\proj{n}$ is the projector onto the $n^{th}$ component of the spin-flavor baryon multiplet, and:
\bea
\cal{I}(Q,M_\pi)&=&\frac{1}{d-1}\int \widetilde{d^d k}\;\frac{\vec k^2}{(k^0-Q+i\eps)(k^2-M_\pi^2+i\eps)}\nonumber\\
&=&
\frac{i}{3(4\pi)^2}\bigg( Q\Big((3M_\pi^2-2Q^2)\Big(\lambda_\eps-\log \frac{M_\pi^2}{\mu^2}\Big)+7M_\pi^2-\frac{16}{3}Q^2\Big)\nonumber\\
&+&  4(M_\pi^2-Q^2-i\eps)^{3/2}\Big(\frac \pi 2- \arctan\frac{Q}{\sqrt{M_\pi^2-Q^2-i\eps}}\Big)\bigg)\nonumber\\
\cal{I}(M_\pi)&=& \int \widetilde{d^d k}\; \frac{1}{k^2-M_\pi^2+i\eps}=\frac{i}{(4\pi)^2} M_\pi^2 \Big(\lambda_\eps +1-\log\frac{M_\pi^2}{\mu^2}\Big) ,
\eea
\end{widetext}
where $d=4-2\eps$, $\widetilde{d^d k}=\frac{d^d k}{(2\pi)^d}$, and $\lambda_\eps\equiv \frac{1}{\eps} -\gamma_E+\log 4\pi$.
 
The contribution to the wave function renormalization factor is given by:
\beq
\delta Z(p^0)=-\frac{\partial}{\partial p^0}\Sigma(p^0).
\eeq

For a particular baryon state $n$, the one-loop contribution to the mass and the wave function renormalization is evaluated by replacing $p^0\to \delta m_n$. With this, the UV divergent pieces can be expressed as spin-flavor operators for the mass shift and wave function renormalization, namely:
\begin{widetext}
\bea
\Sigma^{UV}&=& -\frac{\lambda_\eps}{(4\pi)^2} \left(\left(\frac {\gA}{F_\pi}\right)^2\Big(M_\pi^2 G^{ia}[\delta\hat m,G^{ia}]-\frac 23 G^{ia}[\delta\hat m,[\delta\hat m,[\delta\hat m,G^{ia}]]]\Big)+2 c_1 \frac{N_c M_\pi^4}{F_\pi^2 \Lambda}\right) \nonumber\\
&=&\frac{\lambda_\eps}{(4\pi)^2} \left(\left(\frac {\gA}{F_\pi}\right)^2\frac{C_{\rm HF}}{N_c}\Big(M_\pi^2(-\frac 38 N_c(N_c+4)+\frac 52 \hat S^2)\right. \nonumber\\
&+&\left.  \frac 13 \left(\frac{C_{\rm HF}}{N_c}\right)^2(3N_c(N_c+4)+(5N_c(N_c+4)-24)\hat S^2-28 \hat S^4)\Big)+2 c_1 \frac{N_c M_\pi^4}{F_\pi^2 \Lambda}\right)\nonumber\\
\delta Z^{UV}&=&-\frac{\lambda_\eps}{(4\pi)^2}\left(\frac {\gA}{F_\pi}\right)^2(M_\pi^2 \hat G^2-2G^{ia}[\delta\hat m,[\delta\hat m,G^{ia}]])\nonumber\\
&=&\frac{\lambda_\eps}{(4\pi)^2}\left(\frac {\gA}{F_\pi}\right)^2\Big(\frac{1}{16}M_\pi^2(-3 N_c(N_c+4)+8 \hat S^2) \nonumber\\
&+&  \frac 12 \left(\frac{C_{\rm HF}}{N_c}\right)^2 (3N_c(N_c+4)+2(N_c(N_c+4)-12)\hat S^2-8 \hat S^4)\Big).
\eea
\end{widetext}
Note that the UV divergent piece of the mass is $\ord{N_c^0}$ and driven by the mass splitting term. On the other hand the non-analytic contributions to the mass are $\ord{N_c}$. In the large $N_c$ limit, the wave function renormalization correction is proportional to $\ord{N_c}\times M_\pi^2$: this indicates that approaching the large $N_c$ limit with a baryon effective theory based on perturbation theory is not consistent. This is one additional argument in favor of the linked $\xi$-expansion.

The baryon mass  differences remain as $\ord{1/N_c}$, and for $N_c=3$ one obtains:
\begin{widetext}
\bea
m_\Delta-m_N&=&C_{\rm HF}+\left(\frac{\gA}{F_\pi}\right)^2 \frac{1}{(4 \pi)^2 }  \left(\frac 16 C_{\rm HF} \Big(5\left(3{M_\pi^2}-2 {C_{\rm HF}^2} \right)  (\lambda_\eps -\log  \frac{M_\pi^2}{\mu ^2}  )+35 M_\pi^2-\frac{80
	C_{\rm HF}^2}{3}\Big)\right.\nonumber\\
		&+& \left(M_\pi^2-C_{\rm HF}^2\right)^{3/2} \Big(\pi
	-\frac{10}{3} \arctan \frac{C_{\rm HF}}{\sqrt{M_\pi^2-C_{\rm HF}^2}} \Big)\bigg)+CT,
	\label{massdiff}
\eea
\end{widetext}
where $CT$ indicates the mass counterterms.
 \subsection{Axial   and $\pi B B'$ couplings}
 
 At LO the axial vector current matrix elements and pion-baryon interactions are given by:
 \bea
 \langle B'|A^{ia}\mid B\rangle&=&\gA \langle B'|G^{ia}\mid B\rangle\nonumber\\
 \langle B'|{\cal{L}}_B^{(1)}\mid\! \pi^a(k) B\rangle&=& -i \frac{\gA}{F_\pi} k^i \langle B'|G^{ia}\mid\!\! B\rangle,
 \eea
 satisfying the Goldberger-Treiman relation.
 
The NNLO  one-loop contributions are shown in Fig.(\ref{Axial-currents-loop}). The analysis of  the axial currents  was presented in  Ref. \cite{CalleCordon:2012xz}.

  
 The $\pi B B'$ vertex to one-loop order is given by:  
\begin{widetext}
 \bea
 i\Gamma^a(k,p,p')&=& \frac{\gA}{F_\pi} k^i\left  (\phantom{\frac 1 1}\!\!\! G^{ia}
 +i \left(\frac {\gA}{F_\pi}\right)^2 \sum_{n,n'}  
 \int \widetilde {d^dk'} \frac{G^{jb}\proj{n' }G^{ia} \proj{n} G^{lb}\;\,k'^jk'^l}{(k'^2-M_\pi^2)(p^0-k'^0-\delta m_n)(p'^0-k'^0-\delta m_{n'})}\right. \nonumber\\
 &-& \left. \frac 12 \{\delta Z,G^{ia}\}-\frac 13 \frac{1}{F_\pi^2} G^{ia} {\cal{I}}(M_\pi) 
 + {\rm CTs}\right), \nonumber\\
 &=& \frac{\gA}{F_\pi} k^i\left  (G^{ia}
 -i \bigg(\frac {\gA}{F_\pi}\right)^2 \sum_{n,n'} G^{jb}\proj{n' }G^{ia} \proj{n} G^{jb}\frac{{\cal{I}}(\delta m_n-p^0,M_\pi)-{\cal{I}}(\delta m_{n'}-p'^0,M_\pi)}{p^0-p'^0-\delta m_n+\delta m_{n'}} 
 \nonumber\\
 &-&   \frac 12 \{\delta Z,G^{ia}\}-\frac 13 \frac{1}{F_\pi^2} G^{ia} {\cal{I}}(M_\pi) 
 + {\rm CTs}\bigg),
 \eea
 \end{widetext}
 where the   term involving the wave function renormalization correction is key to restoring the $N_c$ consistency,   canceling contributions $\ord{N_c^{3/2}}$ from  the first diagram (A). 
 
\begin{widetext}
 \begin{center}
 	\begin{figure}[b!]
 		\centerline{\includegraphics[width=14.cm,angle=-0]{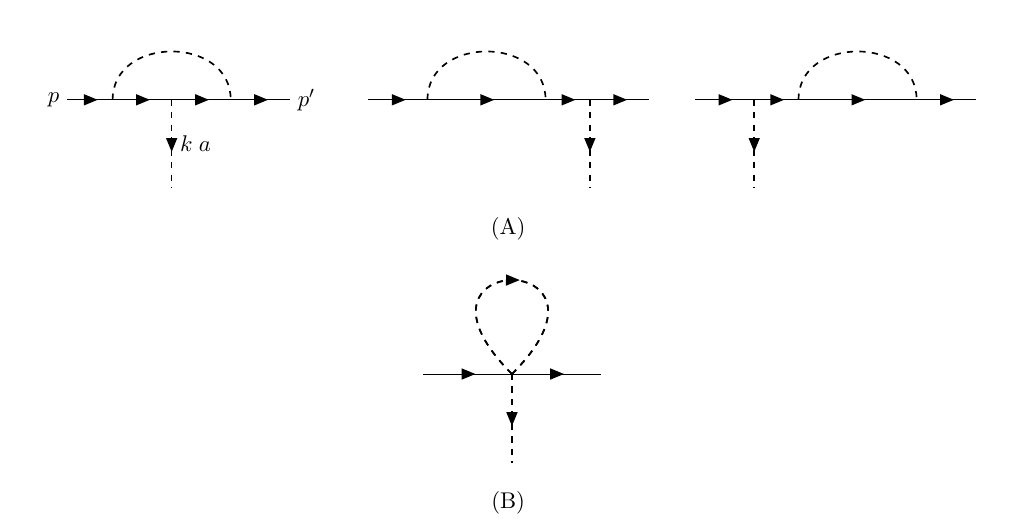}}
 		\caption{One-loop diagrams for the pion-baryon interaction.}
 		\label{Axial-currents-loop}
 	\end{figure}
 \end{center}
\end{widetext}
 
 For baryon momenta on-shell, i.e. setting $p^0=\delta m_{in}$, $p'^0=\delta m_{out}$,  the UV divergence can be given in the operator form:
 \begin{widetext}
 \bea
 i\Gamma^a_{UV}&=&-\frac{\lambda_\eps}{(4\pi)^2}\frac{\gA}{F_\pi}k^i\bigg(-\frac{M_\pi^2}{F_\pi^2} G^{ia}+\frac 16 \left( \frac{\gA}{F_\pi}\right)^2\Big( \big(2\left(\frac{C_{\rm HF}}{N_c}\right)^2(N_c(N_c+4)-2)-3 M_\pi^2\big)G^{ia}   \nonumber\\
 &+&   10 \left(\frac{C_{\rm HF}}{N_c}\right)^2(\{\hat S^2,G^{ia}\}-(N_c+2) S^i I^a)\Big)\bigg).
 \eea
 \end{widetext}
 The UV divergent pieces illustrate two most important facts about the higher order corrections to the pion-baryon couplings: i) the quark mass contributions are suppressed as $\ord{1/N_c}$ with respect to the LO and do not affect the spin-flavor structure of the LO, and ii) the terms that affect the spin-flavor structure, i.e. terms proportional to $S^i I^a$, are corrections of $\ord{1/N_c^2}$ with respect to the LO. Finally, the Goldberger-Treiman discrepancy remains, as it is well known, an effect that is only result of a CT. For the CT Lagrangian see Appendix \ref{app:Lagrangians} Eqn.(\ref{eq:axial-current-CT-lagrangian}). The LECs cannot be fixed completely as at $N_c=3$ the only experimentally accessible   couplings are  $g_{\pi NN}$, $g_{\pi N\Delta}$, and the $g_A$ of the nucleon. In the present analysis the $\pi N\to \pi N$ amplitudes will be used to determine them.

\section{  Scattering amplitudes }  \label{sec:scattering-amplitudes}

This section presents the formalism for the general case of $\pi B\to \pi B'$ scattering. It is presented for general $N_c$ for the purpose of identifying the order in $N_c$ of the different contributions. A rigorous implementation of renormalization at the NNLO is carried out.

\begin{widetext}
	\begin{center}
		\begin{figure}[hbt!]
			\centerline{\includegraphics[width=14.cm,angle=-0]{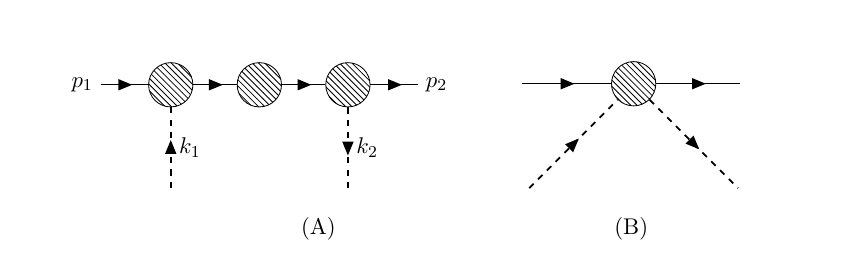}}
			\caption{General decomposition of the scattering amplitude into pole (A) and no-pole (B) contributions. To diagram $(\rm A )$ the crossed diagram must be added.}
			\label{Fig-pole-nopole-diagrams}
		\end{figure}
	\end{center}
\end{widetext}

The contributions to the scattering amplitude are decomposed into those with baryon pole singularities and those without, as depicted in Fig.(\ref{Fig-pole-nopole-diagrams}). In the one-loop diagrams,  this decomposition is performed as explained in Appendix \ref{app:NNLO-amplitudes}.

\subsection{LO  scattering amplitude}
The LO T-matrix, given by the diagrams in Fig.(\ref{LO-Amplitude}) and expressed as an operator in spin-flavor reads:

\begin{widetext}
\bea
i  T^{ba}_{LO}&=& -i \left(\frac{\gA}{F_\pi}\right)^2 k_1^i k_2^j \sum_{n} \Big(  \frac{G^{jb} \proj{n} G^{ia}}{p_1^0+k_1^0-\delta m_n+i\eps}+  \frac{G^{ia} \proj{n} G^{jb}}{p_1^0-k_2^0-\delta m_n+i\eps}  \Big)\nonumber\\
&+& \frac{1}{F_\pi^2}   \Big(-2ic_1 N_c \frac{M_\pi^2}{\Lambda}\delta^{ab}+\frac 12 (k_1^0+k_2^0) \eps^{abc}I^c \Big),
\label{LOamplitude}
\eea

\begin{center}
\begin{figure}[h!] 
\centerline{\includegraphics[width=12.cm,angle=-0]{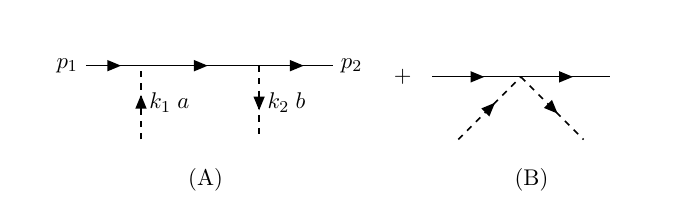}}
	\caption{LO contributions to the scattering amplitude. To diagram $(\rm A) $ the crossed diagram must be added.}
	\label{LO-Amplitude}
\end{figure}
\end{center}
\end{widetext}
  
 This amplitude is precisely the one that requires the emergence of spin-flavor symmetry \cite{Dashen:1993as}, since the individual terms proportional to $\gA^2$ have contributions  $\ord{N_c}$,   and such contributions must cancel when the terms are added. Moreover, the pole terms reflect the non-commutative nature of the chiral and $1/N_c$ expansions. The LO pole terms will combine with the higher order ones to yield the result shown in Eqn.(\ref{NNLOpoleamplitude}).

 \subsection{One-loop corrections to the $\pi B$ scattering amplitude}
 
 The scattering amplitudes at NNLO involve the one-loop diagrams of Figs.(\ref{Fig-loops-gA4},\ref{Fig-loops-gA2},\ref{Fig-loops-gA0}) and the CT contributions from Lagrangians Eqns.(\ref{UV-CTs4},\ref{UV-CTs2},\ref{UV-CTs0},\ref{Finite-CTs}), which provide for the renormalization. The loops are organized in three groups according to their power in $\gA$.  
 
 \begin{widetext}
 	\begin{center}
 		\begin{figure}[hbt!]
 			\centerline{\includegraphics[width=12.5 cm,angle=-0]{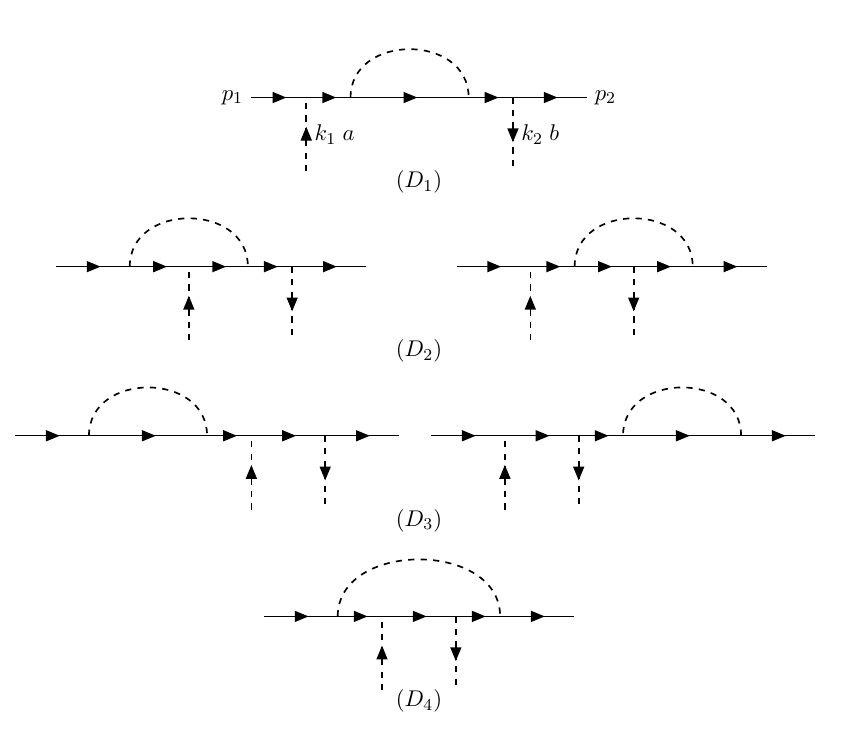}}	
 			\caption{Diagrams proportional to $\gA^4$}
 			\label{Fig-loops-gA4}
 		\end{figure}
 	\end{center}
 	\begin{center}
 		\begin{figure}[t!]
 			\centerline{\includegraphics[width=12.cm,angle=-0]{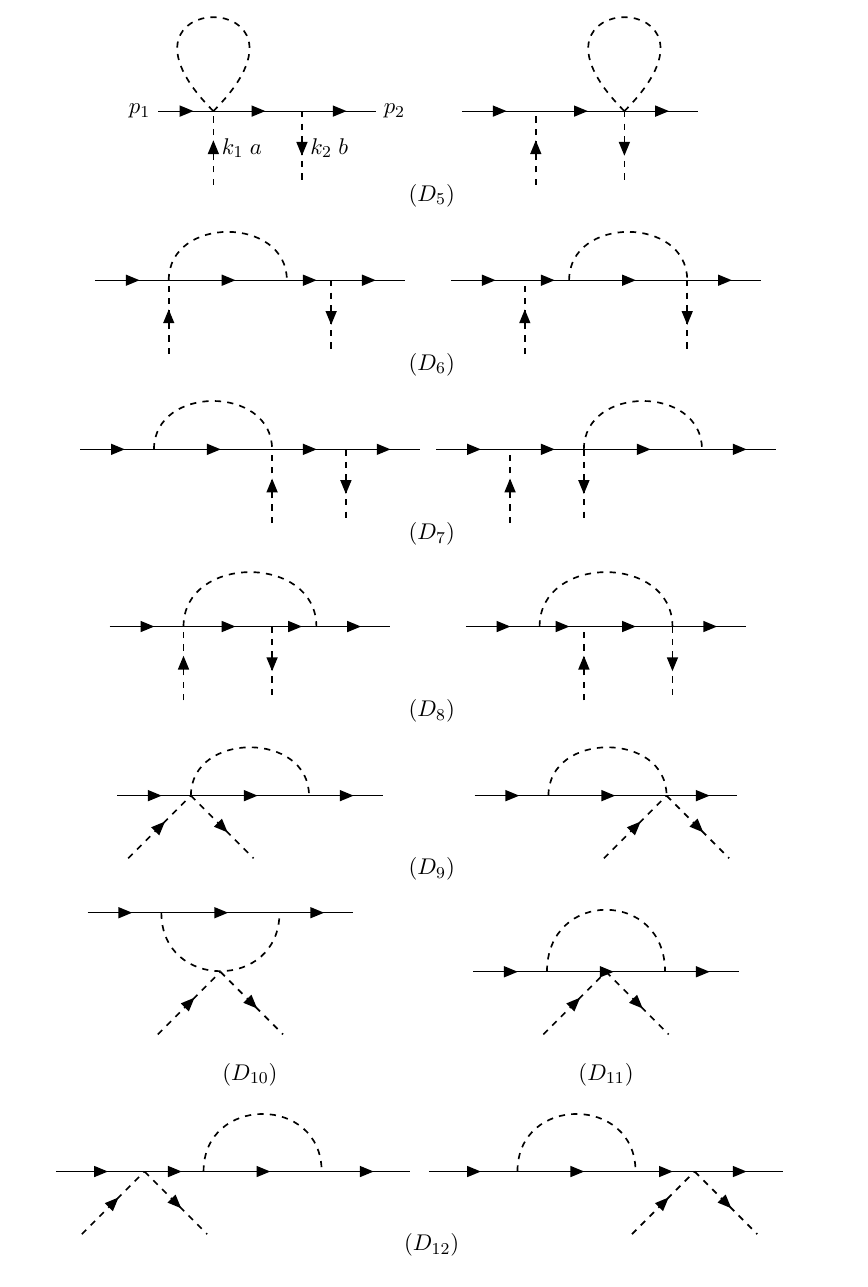}}
 			\caption{Diagrams proportional to $\gA^2$. Diagrams $D_{6,7,8}$ vanish identically. Tadpole diagrams, which only renormalize the baryon mass, are not shown.}
 			\label{Fig-loops-gA2}
 		\end{figure}
 	\end{center}
 \end{widetext}
 
  For one- and two-particle reducible diagrams, the decomposition into pole and no-pole terms is carried out following Eqns.(\ref{apole},\ref{bpole}). The results for the individual diagrams, including   details on the UV divergencies, are presented in Appendix \ref{app:NNLO-amplitudes}. 
 The spin-flavor tensors in the diagrams are projected in t-channel $(J,I)$, where  $J,\;I=0,1,2$. Since the ground state baryon spin-flavor multiplet consists only of states $S=I$,   only transitions with baryon $\Delta S=0,1,2$ can occur. In the physical case $N_c=3$, $\Delta S=0,1$.    The CT Lagrangians for renormalizing the no-pole contributions are then constructed following those t-channel projections, using the basis of spin-flavor operators in Appendix \ref{app:SFOperators} Table \ref{SF-basis}. Those Lagrangians contain terms $\ord{\xi^2\; \&\; \xi^3}$. Throughout, in the construction of the higher order Lagrangians the LO equations of motion are used, namely:
 
 \bea
 i D_0 \B&=&(\frac{C_{\rm HF}}{N_c}-c_1 \frac{N_c}{2\Lambda}\langle \chi_+\rangle-\gA u^{ia} G^{ia})\B\nonumber\\
 D_\mu u^\mu&=& \frac i 2 \chi_-,
 \eea
along with the identities:
 \bea
 D_\mu u_\nu-D_\nu u_\mu& = & -f_{-\mu\nu}\nonumber\\
 \,[D_\mu,D_\nu\,]&=& -i \Gamma_{\mu\nu}\nonumber\\
 \Gamma_{\mu\nu}&=& \frac 12 f_{+\mu\nu}+\frac 14[u_\mu,u_\nu].
 \eea
 
\begin{widetext}
\begin{center}
\begin{figure}[hbt!]
	\centerline{\includegraphics[width=9.cm,angle=-0]{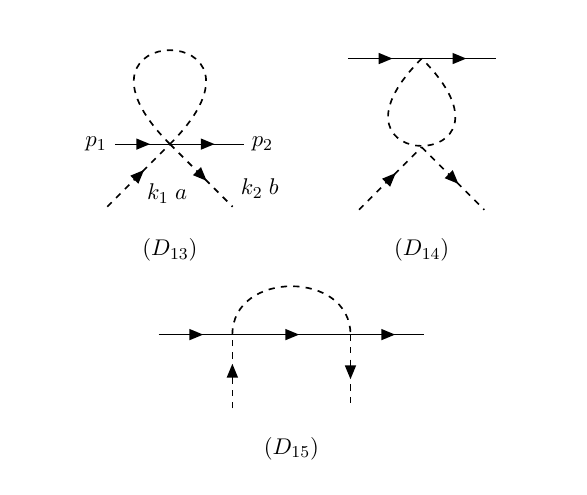}}
\caption{Diagrams proportional to $\gA^0$.}
\label{Fig-loops-gA0}
\end{figure}
\end{center}
\end{widetext}

For each of the sets of diagrams, the sum of the no-pole UV divergent parts is decomposed into t-channel $(J,I)$ and, for general $N_c$, the corresponding spin-flavor tensors are expressed  in terms of the basis in Table \ref{SF-basis}. In the set proportional to $\gA^4$,  individual diagrams are $\ord{N_c^2}$, thus two orders must exactly cancel in the sum of the UV divergent pieces, which is indeed the case as shown in the result Eqns.(\ref{UVgA4},\ref{UVgA4red},\ref{pUVgA4}). Similarly, the set proportional to $\gA^2$, where diagrams $D_{6,7,8}$ vanish identically, has diagrams $\ord{N_c}$, with such contributions cancelling in the sum of the UV divergent pieces, as shown by Eqn.(\ref{UVgA2}).  Finally, the diagrams proportional to $\gA^0$ are individually $\ord{N_c^0}$, and have the property  that they involve only a fixed spin baryon state throughout the diagrams. 

In the CT Lagrangians, the LECs are expressed in the   form $X=\beta_X \frac{\lambda\eps}{(4\pi)^2}+X(\mu)$, where $X(\mu)$ is the renormalized LEC at the scale $\mu$, and the $\beta$ function  $\beta_X$ is  adjusted to eliminate the corresponding UV divergence  in the $\overline{MS}$ subtraction scheme. The   $\beta$ functions are shown in Table \ref{betafunctionstable}.
There are  several interesting observations to be made, namely: i) The T-matrix UV divergencies proportional to $\gA^4$ have no $M_\pi$ dependencies. ii) The UV divergencies contain different orders in $\xi$, as shown in Eqns.(\ref{UVgA4},\ref{UVgA2},\ref{UVgA0}). iii) In all cases the renormalization scale $\mu$ appears in the single combination $\log\frac{M_\pi^2}{\mu^2}$. iv) All non-analytic terms involving the residual masses of the baryons, i.e., $\delta m_n$, appear in combination with the pion mass in factors of the  form $ (\delta m_n-\delta m_{n'})^\nu\log\frac{M_\pi^2}{\mu^2}$ $ (\nu=1,3)$, $(M_\pi^2 - (\delta m_n-\delta m_{n'})^2)^{\nu/2}$  $(\nu=1,3)$,   $\arctan\frac{(\delta m_n-\delta m_{n'})}{\sqrt{M_\pi^2 - (\delta m_n-\delta m_{n'})^2}}$. v) The evaluations of the matrix elements of the spin-flavor tensors for the non-analytic contributions, which involve projections on different baryon intermediate states, are performed explicitly using the results for the different matrix elements of the $SU(4)$ generators in Eqn.(\ref{ME-generators}), as the results cannot be  reduced to simple expressions in terms of the basis spin-flavor tensors.

Adding the pole terms from the leading-order amplitude, Eqn.(\ref{LOamplitude}), together with the higher-order pole contributions from the diagrams $D_1$, $D_2$, $D_3$, and $D_5$, and after performing the mass and pion–baryon coupling renormalizations,  results in the following general form of the pole amplitude: 
\begin{widetext}
\bea
\langle B'\mid i T^{ba}_{\rm pole}\mid B\rangle&=&-i \frac{k_1^i k_2^j}{F_\pi^2}\; \sum_{n}\mathring g_{ABB_n}\;\mathring g_{AB_n B'} \left(\frac{\langle B'\mid G^{jb}\proj{n}G^{ia}\mid\!\! B\rangle}{p_B^0+k_1^0-\delta m_n^R+\frac i 2 \Gamma_n}+\text{ crossed}
\right),
\label{NNLOpoleamplitude}
\eea 
\end{widetext}
where $\mathring g_{ABB'}$ is the renormalized coupling, which at LO is equal to $\gA$, $\delta m_n^R$ is the renormalized residual mass, $p^0_B=\delta m_B^R+\frac{\vec k_1^2}{2 m_0}$ is the residual energy of the external baryon that includes the kinetic energy   $\ord{\xi^3}$, and $\Gamma_n=\ord{\xi^2}$ is the decay width. Such terms must be included at the NNLO of the calculation. At $N_c=3$, only $\Gamma_\Delta$ is needed and reads:
 \beq
\Gamma_\Delta=\frac{\mathring g_{AN\Delta}^2 m_N}{6\pi (16 F_\pi)^2 m_\Delta^4} \lambda_K(m_\Delta^2,m_N^2,M_\pi^2)^{3/2},
\label{Deltawidth}
\eeq
 where $\lambda_K$ is the K\"allen phase space function. Because of the P-wave nature of the decay, in order to be accurate enough, the exact phase space factor must be used.


\section{$\pi N \to \pi N$  scattering and fit to data.}\label{sec:piN-piN-fits}

This section aims at testing the effective theory by fitting to the $\pi N \to \pi N$ data, for which the SAID data base \cite{Arndt:2006bf,SAID-webpage} is used.

The T-matrix is expressed in the standard CM t-channel decomposition  in terms of the spin-non-flip $g^\pm$ and spin-flip $h^\pm$ amplitudes:
\begin{widetext}
\bea
T^{ba}_{\pi N\to \pi N}&=& \frac{E_N+m_N}{2m_N}\left( \delta^{ab}(g^+(s,t)-i (\vec k_1\times \vec k_2)\cdot \vec \sigma \;h^+(s,t))\right.\nonumber\\
&+& \left. i\eps^{bac}\tau^c(g^-(s,t)-i (\vec k_1\times \vec k_2)\cdot \vec \sigma \;h^-(s,t))\right),
\eea
giving the s-channel partial wave amplitudes:
\bea
f^{\pm}_{\ell\pm}(s)&=& \frac{E_N+m_N}{16 \pi \sqrt{s}}\int_{-1}^{1}dz\left(g^\pm(s,t(z))P_\ell(z))+k^2 h^\pm(s,t(z))(P_{\ell\pm 1}(z)-z P_\ell(z)\right),
\eea
\end{widetext}
where $t(z)=-2k^2(1-z)$, $k$ is the CM momentum, and $z=\cos\theta$,  $\theta$ being the CM  scattering angle. The projection to s-channel isospin is given by the well known relation:
\beq
\left(\begin{array} {c}  f_{\ell\pm}^{1/2}\\
f_{\ell\pm}^{3/2}\end{array}
\right)=\left(\begin{array} {cc}1&2\\1&-1 \end{array} \right)\left(\begin{array} {c}  f_{\ell\pm}^{+}\\
	f_{\ell\pm}^{-}\end{array} \right).
\eeq
Throughout the exact kinematics for $E_N$, $s$ and $t$ are used.

For a generic partial wave amplitude the corresponding phase shift is defined by:
\beq
f(s)=-\frac{i}{2k}(e^{2i\delta(s)} \eta(s)-1),
\label{unitarity-relation}
\eeq
with the inelasticity factor $0\leq\eta(s)\leq 1$. In the present case, where the analysis is carried out below the onset of the second resonance region, it turns out that   one can approximate $\eta\sim 1$ throughout  \cite{Arndt:2006bf,SAID-webpage}. Unitarity is only approximate in the effective theory, and only ${\rm Re} f(s)$ is affected by the NNLO LECs to be fitted, while ${\rm Im} f(s)$ is entirely determined by the LO Lagrangian, and is thus being less accurate than the real part. It is therefore more realistic to fit to the real part of the partial wave amplitudes and then use Eqn.(\ref{unitarity-relation}) to give the imaginary part through the obtained phase shift. Only a subset of the NNLO LECs in the CT  Lagrangians can be determined by $\pi N\to \pi N$ scattering. Those are shown by the CT contributions to the T-matrix in Eqn.(\ref{piNCTR}).

Fits to the SAID single-energy solutions were performed in different ranges of pion CM momentum, up to a maximum of 350 MeV, which corresponds to $\sqrt{s}=1.38$ GeV. In the fits, the   S-wave scattering lengths \cite{Schroder:2001rc,Hirtl:2021zqf} are inputs. The real parts of the amplitudes up to and including D-waves along with the imaginary part of the  $P_{33}$ amplitude are fitted. The latter presents an imaginary part that in the case of the pole contribution involves the NNLO couplings and masses,  is thus accurate to the NNLO, and ought to be   included in the fit. The results of the fit up to $k=350$ MeV along with the  phase shifts are shown in Figs.(\ref{Fig-partial-waves},\ref{Fig-phase-shifts}). Depicted are the imaginary parts obtained via unitarization from the real parts, as well as the imaginary parts that result from the perturbative calculation, which in general are less accurately  described for the reason mentioned earlier.

In the range $k<400$ MeV the inelasticity factor is taken to be unity for all partial waves, as it can be confirmed by the SAID analysis \cite{Arndt:2006bf,SAID-webpage}. In particular this requires for the $P_{33}$ partial wave that  $k\times {\rm Re} f_{1+}^3$  reaches a maximum value $+1/2$ and a minimum value $-1/2$ around the $\Delta$ resonance, which is fulfilled by the fit within the error band.  
Note that the $P_{33}$ pole contributions have the form of the Breit-Wigner for the $\Delta$ pole. The no-pole terms from the non-analytic and CT contributions are essential for restoring consistency with the unitarity constraint on ${\rm Re} f_{1+}^3$ Eqn.(\ref{unitarity-relation}). 
The errors of the inputs are in general much smaller than the estimated theoretical error of the NNLO calculations. The relative theoretical error is $\ord{\xi^3}$, which is likely to be much larger than the  percent errors of the most precise inputs, namely those of S- and P-waves. The rather large $\chi^2$ per degree of freedom of the fits is indicative of that disparity. 
An   error band for the theoretical amplitudes is obtained  by performing 100 bootstrap resamplings of the data for each partial wave.  The parameters resulting from the fit are shown in Table \ref{fitparameters}.
\begin{table}[h!] 
\begin{tabular}{l c|cr}
 \hline\hline
$	C_{\rm HF}[MeV] $ & $ 304.49(  1.16)$ & $	\;\; \alpha_{00}^{(1)} $ & $ 8.55  ( 0.21)$ \\ $
\Gamma_\Delta [MeV] $ & $ 119.5( 1.45)$ & $	\;\; \alpha_{00}^{(2)} $ & $ -5.56 (  0.19)$ \\ $
	\gA $ & $ 1.18 (0.)$ & $	\;\; \alpha_{00}^{(4)} $ & $ -0.48 ( 0.17)$ \\ $
	c_1 $ & $ 0.46 ( 0.) $ & $	\;\; \alpha_{01}^{(1)} $ & $ 3.82 ( 0.49)$ \\ $
	g_{ANN} $ & $ 1.59 ( 0.03) $ & $	\;\; \alpha_{01}^{(2)} $ & $ -0.88 ( 0.66)$ \\ $
g_{AN\Delta} $ & $ 1.81  (  0.01)$ & $	\;\; \alpha_{01}^{(3)} $ & $ 1.01 ( 0.15)$ \\  
& & $	\;\; \alpha_{01}^{(5)} $ & $ -0.98 ( 0.61)$ \\ 
&&$	\;\; \alpha_{10}^{(1)} $ & $ 2.82 ( 0.57)$ \\ 
&&$	\;\; \alpha_{11}^{(1)} $ & $ -6.25 ( 0.84)$ \\ 
&&$	\;\; \alpha_{11}^{(2)} $ & $ 4.53 ( 2.26)$ \\ 
	\hline\hline
\end{tabular}
\caption{Low energy constants from fit. The range for the fit parameters   in brackets indicate the range of variation obtained via bootstrap resampling, and should not be confused with actual uncertainty. The results correspond to the choice  $\Lambda=\mu=m_\rho$.}
\label{fitparameters}
\end{table}
\%end{widetext}

This analysis of the partial wave amplitudes gives strong support for the implementation of the $1/N_c$ consistency conditions in the framework of $\rm{BChPT \times 1/N_c}$ using the $\xi$-expansion. A key observation is the range of validity of the expansion: at NNLO, the approach yields a consistent description of the amplitudes for center-of-mass (CM) pion momenta up to 200–350 MeV, depending on the specific partial wave.

The real parts of the amplitudes are well described within different ranges of CM pion momentum, namely: the S-waves are reliably reproduced up to about 250 MeV; the P-waves show good agreement up to 250–350 MeV; the D-waves are more challenging due to their naturally small magnitudes, relatively large uncertainties, and lower sensitivity to contact term (CT) contributions. Among the D-wave channels, the fits to $D_{13}$ and $D_{35}$ are consistent with expectations, while those for $D_{33}$ and $D_{15}$ exhibit discrepancies that merit further investigation.

A  comparison with previous BChPT analyses that include explicit $\Delta$ degrees of freedom \cite{Fettes:2000bb,Torikoshi:2002bt,Alarcon:2011zs} indicates that the $\rm{BChPT \times 1/N_c}$ framework achieves a comparable, and in some cases even broader, energy range of agreement with experimental data.




The result  for the  coupling   $\mathring g_{ANN}$, disregarding the Goldberger-Treiman discrepancy, implies a determination of the nucleon axial coupling $g_A=1.325$, and the result for the $\mathring g_{AN\Delta}$ coupling gives $\Gamma_\Delta= 134$ MeV, which is about 10\% larger than the fitted value. Note that the fitted value agrees well with the Breit-Wigner width's estimate from the PDG \cite{ParticleDataGroup:2024cfk}. 

The fit result for  $C_{\rm HF}$ is similar to the LO one if one takes for the $\Delta$ mass the Breit-Wigner PDG estimate. On the other hand, the LO $\gA=1.52$, obtained from the nucleon axial coupling, is reduced by about 20\% at the NNLO. Finally, the S-wave scattering lengths resulting from the fit are $a^+=0.012/M_\pi$ and $a^-=-0.087/M_\pi$, to be compared with the  experimental ones  \cite{Schroder:2001rc,Hirtl:2021zqf} $a^+=(0.0078\pm 0.0028)/M_\pi$ and $a^-=(0.0866\pm 0.0010)/M_\pi$.

\begin{widetext}
	\begin{center}
		\begin{figure}[hbt!]	
			\includegraphics[width=0.65\columnwidth]{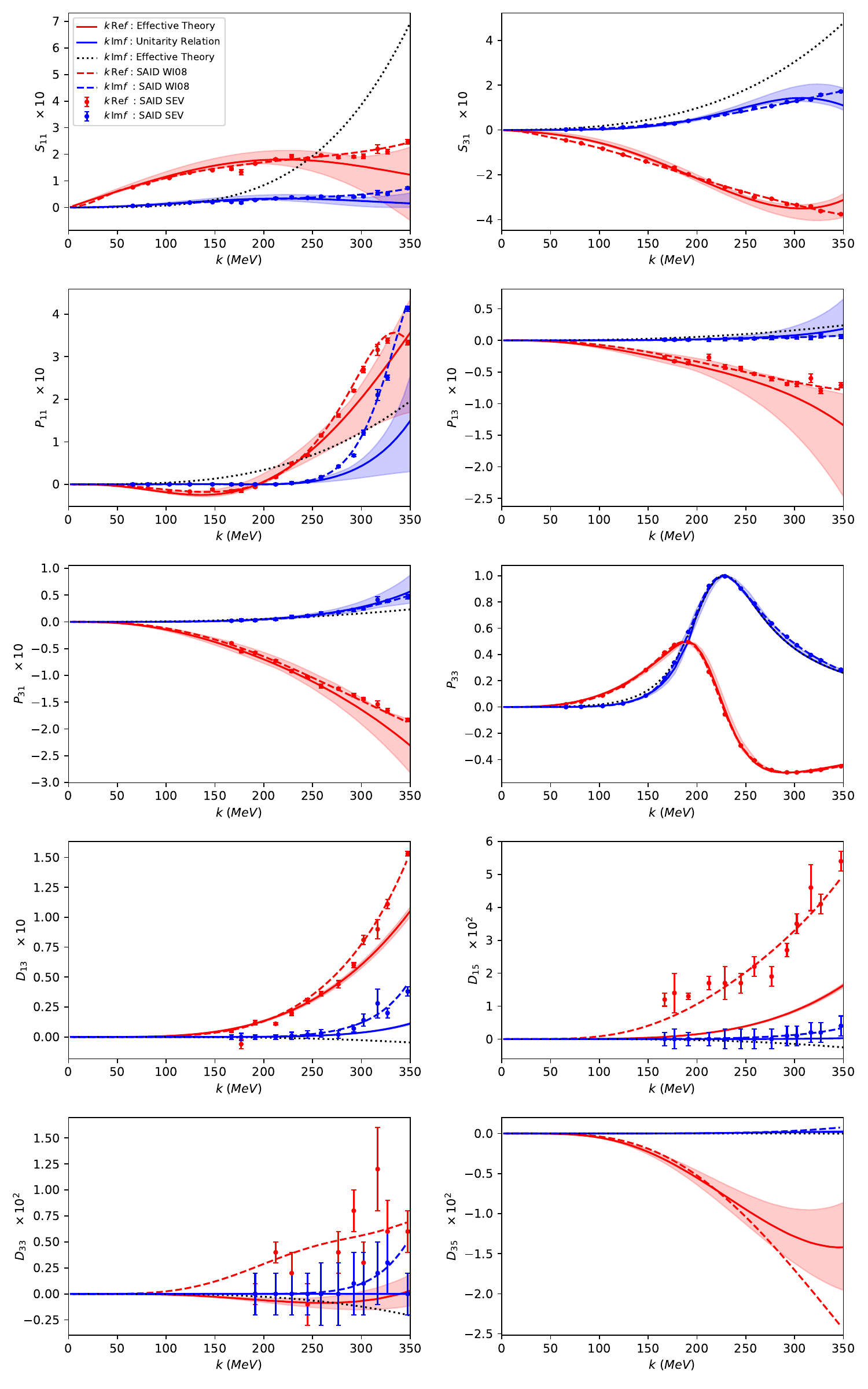}
			\caption[]{Fits to   the partial wave amplitudes $ {\rm Re} f$  and the  $ {\rm Im} f$ of the $P_{33}$ partial wave. Plots show $ k \times{\rm Re} f$ (solid red) and $ k \times {\rm Im} f$ vs $k$ from phase shift as determined from ${\rm Re} f$ using the unitarity-relation Eqn.(\ref{unitarity-relation}) (solid blue), and from absorptive parts of one-loop diagrams (dotted black). The error band estimates are obtained by bootstrap resampling the data. The data is from the SAID data base \cite{Arndt:2006bf,SAID-webpage}: WI08 solution (dashed), and the   data points, used for the fits, are the single energy solution SEV.  }
			\label{Fig-partial-waves}
		\end{figure}
	\end{center}
	\begin{center}
		\begin{figure}[hbt!]	
			\includegraphics[width=0.650\columnwidth]{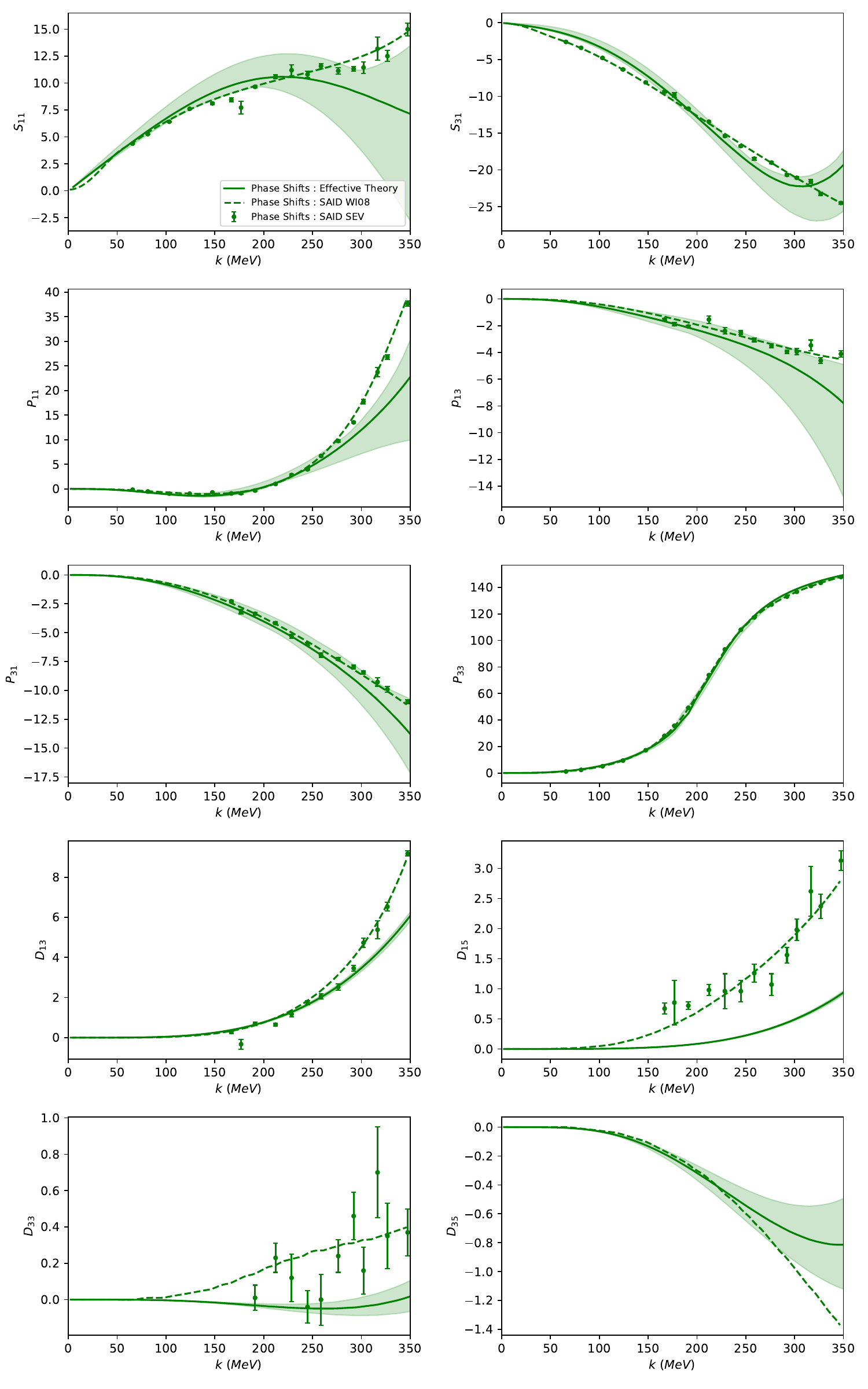}
			\caption[]{Phase shifts: SAID WI08 solution (dashed),data points from SEV solution, and   extracted from fit in Fig.(\ref{Fig-partial-waves}) using unitarity relation  Eqn.(\ref{unitarity-relation}) (solid).}
			\label{Fig-phase-shifts}
		\end{figure}
	\end{center}
\end{widetext}

\section{Summary and Conclusions} \label{sec:conclusions}

This work implements the combined framework of BChPT and the $1/N_c$ expansion for low-energy pion–baryon scattering at next-to-next-to-leading order (NNLO) in the $\xi$-expansion. This approach respects both the constraints of chiral symmetry and the systematic expansion in $1/N_c$ derived from QCD. The $\xi$-expansion retains, without further expansion, non-analytic terms where the chiral and $1/N_c$ expansions are not simultaneously valid—terms that typically exhibit slow convergence in either expansion performed individually. This treatment is essential for realistic descriptions of QCD in the physical world. The role of the $SU(4)$ spin–flavor symmetry is central to ensuring manifest consistency with the large-$N_c$ limit of observables. In particular, it leads to a suppression of loop corrections via cancellations among diagrams that would otherwise violate large-$N_c$ scaling.

A test of the framework with elastic $\pi N$ scattering shows a natural range of applicability for pion CM energy up to about 350 MeV, and somewhat larger for the $P_{33}$ channel. The range of S-waves is primarily limited by the contributions from the diagrams proportional to $\gA^0$, which only involve the nucleons throughout, and are therefore the same as in the case of the ordinary BChPT with only active nucleons. While the real parts of the partial wave amplitudes are well described in significant part due to the available NNLO LECs,   departures  from unitarity are manifested in the imaginary parts, in particular in the S-waves above pion energy of 200-250 MeV. Those departures happen  in amplitudes where those absorptive parts are very small.

 These results motivate further investigations using $BChPT \times 1/N_c$ in the context of low-energy scattering \cite{GoityJayakodige}, particularly focusing on the extraction of the $\pi N$ sigma term and a comprehensive analysis that incorporates currents and a more detailed study of $\pi N$ scattering.
 
 \section*{Acknowledgments}
 This material is based upon work supported by the U.S.~Department of Energy, Office of Science,
 Office of Nuclear Physics under contract DE-AC05-06OR23177 (JLG), and by the National Science
 Foundation, Grant Number PHY 1913562 (JLG, DJ). The authors are indebted Christian Weiss for discussions, and to Ron Workman and Igor Strakovsky for clarifications on the pion-nucleon SAID data base.

\appendix
\section{Chiral building blocks }
\label{app:ChBBs}
Using standard notations, the building blocks needed for  constructing chiral tensors are:
\begin{widetext}
\bea
\Pi(x)&=& \pi^a(x) \tau^a\nonumber\\
u(x)&=& e^{i\frac{\Pi(x)}{2F_\pi}}\;\;\;\;\; U(x)\equiv u^2(x)\nonumber\\
\ell_\mu&=&  v_\mu-a_\mu \;\;\;\;\;\;\;\;\;\; r_\mu=v_\mu+a_\mu\nonumber\\
\Gamma_\mu&=& -\Gamma_\mu^\dag=\frac i2(u^\dag (\partial_\mu-i(v_\mu+a_\mu))u+u(\partial_\mu-i(v_\mu-a_\mu))u^\dag)\nonumber\\
D_\mu&=& \partial_\mu-i \Gamma_\mu\nonumber\\
u_\mu&=&u_\mu^\dag=i(u^\dag (\partial_\mu-ir_\mu)u-u(\partial_\mu-i\ell_\mu)u^\dag)\nonumber\\
\chi&=& 2 B_0(s+i p)\;\;\;\;\;\;  \chi^\dag= 2 B_0(s-i p)\nonumber\\
\chi_\pm&=& \pm \chi_\pm^\dag= u^\dag \chi u^\dag\pm u \chi^\dag u\nonumber\\
\chi_-&=& i \chi_+\arrowvert_{p\leftrightarrow s, u\leftrightarrow u^\dag},
\eea
where $v_\mu$, $a_\mu$ are respectively vector and axial-vector sources, and $s$, $p$ are respectively scalar and pseudoscalar sources.
The local chiral transformations of the building block  are the following:
\bea
(L ,R ): u &=& R  u  h^\dag(L ,R ,u )=h(L ,R ,u )L^\dag  \text{ where } h^\dag h=1\nonumber\\
(L ,R ): \ell_\mu&=&L (\ell_\mu + i \partial_\mu)L^\dag ,\;\;(L ,R ): r_\mu=R (r_\mu + i \partial_\mu)R^\dag \nonumber\\
(L ,R ):u_\mu &=&h(L ,R ,u ) u_\mu  h^\dag(L ,R ,u )\nonumber\\
(L ,R ):\chi_\pm &=&h(L ,R ,u ) \chi_\pm  h^\dag(L ,R ,u ).
\eea

For a matter field, in the present case a baryon, in a given irrep of isospin $SU(2)$, the chiral transformation law is the following:
\beq
(L(x),R(x)):\B(x)=h(L(x),R(x),u(x)) \B(x),
\eeq
\end{widetext}
where the unitary $h$ can be expressed as: 
\beq
h(L,R,u)=e^{i\alpha^a(L,R,u)I^a},
\eeq

with $I^a$ in the given isospin irrep.
For an isospin one operator $X$, its components $X^a$ are given by: $X^a=\frac 12 \langle X \tau^a\rangle$, where $\langle \cdots \rangle$ represents the trace.

\subsection{Expansion of building blocks}
For the purpose of the present work the relevant terms in the building blocks are those with up to four-pion fields, namely:
\begin{widetext}
\bea
u_\mu&=&2 a_\mu-\frac {1}{F_\pi}\partial_\mu\Pi+\frac {i}{F_\pi}[v_\mu,\Pi]+\frac{1}{4 F_\pi^2}[\Pi,[a_\mu,\Pi]]\nonumber\\
&+&\frac{1}{24 F_\pi^3}([\Pi,[\Pi,\partial_\mu\Pi]]-i[\Pi,[\Pi,[v_\mu,\Pi]]])+\cdots\nonumber\\
u_\mu^a&=& \frac 12 \langle u_\mu \tau^a\rangle ~~~\text{ $u_\mu$ in the fundamental irrep}\nonumber\\
\Gamma_\mu&=&v_\mu+\frac{i}{2 F_\pi}[a_\mu,\Pi]+\frac{i}{8 F_\pi^2}[\Pi,[\partial_\mu-iv_\mu,\Pi]] \nonumber\\
&+& \frac{i}{48 F_\pi^3} [\Pi,[\Pi,[\Pi,a_\mu]]]-\frac{i}{384 F_\pi^4} [\Pi,[\Pi,[\Pi,[\partial_\mu-i v_\mu,\Pi]]]]+\cdots\nonumber\\
\chi_+&=& 4 B_0 s+\frac{2B_0}{F_\pi}\{p,\Pi\}-\frac{B_0}{2F_\pi^2}\{\Pi,\{s,\Pi\}\}\nonumber\\
&-& \frac{B_0}{12 F_\pi^3}\{\Pi,\{\Pi,\{ \Pi,p\}\}\}+\frac{B_0}{96 F_\pi^4}\{\Pi,\{\Pi,\{\Pi,\{ \Pi,s\}\}\}\}+\cdots \nonumber\\
\chi_-&=& i(4 B_0 p-\frac{2B_0}{F_\pi}\{s,\Pi\}-\frac{B_0}{2F_\pi^2}\{\Pi,\{p,\Pi\}\}\nonumber\\
&+& \frac{B_0}{12 F_\pi^3}\{\Pi,\{\Pi,\{ \Pi,s\}\}\}+\frac{B_0}{96 F_\pi^4}\{\Pi,\{\Pi,\{\Pi,\{ \Pi,p\}\}\}\}+\cdots), 
\eea
\end{widetext}
where for the present work  the following replacements are made:    $p= 0$, $s={\cal{M}}_q$, where ${\cal{M}}_q$ is the quark mass matrix, and $v^\mu=a^\mu=0$. The following are useful explicit expansions in the pion fields of the building blocks:
\begin{widetext}
\bea
 u_\mu^a&=&-\frac{1}{F_\pi} \partial_\mu \pi^a +\frac{1}{6 F_\pi^3}\pi^b (\pi^b \overleftrightarrow{\partial}_\mu \pi^a) +\cdots\nonumber\\
u_\mu^au_\nu^b&=&\cdots+\frac{1}{F_\pi^2} \partial_\mu \pi^a \partial_\nu \pi^b 
+\cdots\nonumber\\
\langle \chi_+\rangle &=&\cdots+{4} M_\pi^2-\frac{2}{F_\pi^2} M_\pi^2 \pi^a\pi^a+\frac {1}{6F_\pi^4} M_\pi^2  \pi^a \pi^a \pi^b\pi^b+ \cdots\nonumber\\
\chi_+^a &=& \frac{B_0 (m_d-m_u)}{F_\pi^2}(\pi^a\pi^3-2F_\pi^2 \delta^{a3})+\cdots\nonumber\\
\langle \chi_-\rangle &=&\frac{4 i B_0 (m_d - m_u) \pi^3}{F_\pi}+\cdots\nonumber\\
\chi_-^a&=&-\frac{2 i M_\pi^2 \pi^a}{F_\pi}\nonumber\\
D_\mu&=& \partial_\mu+ \frac{i}{2F_\pi^2} \eps^{abc} \pi^a\partial_\mu \pi^b   I^c+\cdots \nonumber\\
D_\mu D^\mu&=&\partial_\mu\partial^\mu+\frac{i}{2F_\pi^2}\eps^{abc }  I^c(\partial_\mu(\pi^a\partial^\mu\pi^b)+2(\pi^a\partial_\mu\pi^b)\partial^\mu+\cdots).
\eea
Additional  bilinears in pion fields needed for the counterterms are the following:
\bea
\langle \chi_+\rangle^2&=& -\frac{16}{F_\pi^2} M_\pi^4 \pi^a\pi^a  \nonumber\\
\langle \chi_+^2\rangle&=& -\frac{2M_\pi^4}{F_\pi^2}  ( \frac{(m_d-m_u)^2}{\hat m^2} \pi^3\pi^3 +4 \pi^a\pi^a)\nonumber\\
\langle \chi_-\rangle^2&=&  -\frac{  16}{F_\pi^2} B_0^2(m_d-m_u)^2 \pi^3\pi^3   \nonumber\\
\langle \chi_-^2\rangle&=& -\frac{2M_\pi^4}{F_\pi^2}  ( \frac{(m_d-m_u)^2}{\hat m^2} \pi^3\pi^3 +4 \pi^a\pi^a)  , 
\eea
\end{widetext}
where, $\hat m=\frac 12 (m_u+m_d)$. Note that in this work, isospin symmetry is assumed.  

\subsection{Discrete symmetry transformations  of building blocks}
The Lagrangians respect the discrete P, C and T symmetries, and the following Table \ref{PCT} provides the necessary transformation rules for the building blocks. Note that C does not apply to the baryon Lagrangians, as the heavy baryon expansion can only describe  either   baryons or   anti-baryons as separate sectors.


Under complex conjugation,   the spin-flavor generators in an arbitrary representation satisfy the relations Eqn. (\ref{commrel}):
\bea
e^{-i \pi S^2}{S^i}^* e^{i \pi S^2}&=&{  -} S^i\nonumber\\
e^{-i \pi I^2} {I^a}^*e^{i \pi I^2} &=&{  -}I^a\nonumber\\
e^{-i \pi S^2}e^{-i \pi I^2}{G^{ia}}^*e^{i \pi I^2}e^{i \pi S^2}&=&G^{ia},
\eea
where $a,i=1,2,3$ and $S^2$ and $I^2$ are the respective generators with  $i,a=2$.

\section{${\mathbf {SU(4)}}$ Algebra}
\label{app:su4}
This Appendix summarizes properties of the $SU(4)$ spin-flavor symmetry group used in the present analysis.
The algebra of $SU(4)$ contains fifteen generators: the spin generators $ {S}^i$, the isospin generators $ {I}^a$,
and the spin-flavor generators $ {G}^{ia}$, where $i$ and $a$ run from 1 to 3. The generator's commutation relations are the following:
\bea
  [  {S}^i, {S}^j ]&=& i \eps^{ijk} {S}^k,~~ [  {I}^a, {I}^b ]=i \eps^{abc}  {I}^c ,\nonumber\\ ~ ~ [  {I}^a, {S}^i ]&=&0,  \nonumber\\    
 	~  [  {S}^i, {G}^{ja} ]&=&i  \eps^{ijk}  {G}^{ka},~~ [ {I}^a, {G}^{ib}]=i \eps^{abc}  {G}^{ic}, \nonumber\\   
 	~   [  {G}^{ia}, {G}^{jb} ]&=& \frac{i}{4}\delta^{ij}\eps^{abc} {I}^c+\frac{i}{4}\delta^{ab} \eps^{ijk}  {S}^k 
 	\label{commrel}
\eea
The ground state baryons  fill the $SU(4)$ totally symmetric irrep corresponding to the Young tableaux with $N_c$ boxes, of dimension $\frac 16(N_c+1)(N_c+2)(N_c+3)$.
These states have spin and isospin $S=I=\frac12,\cdots ,\frac{N_c}{2}$ and are denoted by $|S S_3 I_3\rangle$.
The matrix elements of the $SU(4)$ generators in these states are as follows:
\begin{widetext}
\bea
\langle S' S'_3 I'_3 |  {S}^i | S S_3 I_3\rangle &=& \sqrt{S(S + 1)} \delta_{S'S} \delta_{I_3' I_3}
\langle S S_3, 1 i| S'S_3' \rangle ,
\nonumber\\
\langle S' S'_3 I'_3 |  {I}^a | S S_3 I_3\rangle &=& \sqrt{S(S + 1)} \delta_{S'S} \delta_{I_3' I_3}
\langle S I_3, 1 a| S'I_3' \rangle ,
\nonumber\\
\langle S' S'_3 I'_3 |  {G}^{ia} | S S_3 I_3\rangle&=&\frac 14
\sqrt{\frac{2S+1}{2S'+1}}\sqrt{(N_c+2)^2-(S-S')^2(S+S'+1)^2}\nonumber\\
 &\times&\langle S S_3, 1 i | S' S'_3\rangle\langle S I_3, 1 a | S' I'_3\rangle .
 \label{ME-generators}
\eea
\end{widetext}
For states with $S=\ord{N_c^0}$, $ {S}^i$ and $ {I}^a$ have matrix elements $\mathcal{O}(N_c^0)$ and connect only states with
$S' = S$, while the generators  $ {G}^{ia}$ have matrix elements $\mathcal{O}(N_c)$ and can connect states with
$S' = S$ or $S\pm 1$.

\begin{widetext}
	\begin{center}
		\begin{table}[h!]
			\begin{tabular}{|c|c|c|c|}
				\hline\hline
				&P  &  C& T \\
				\hline
				$x^\mu$ & $x_\mu$ & $x^\mu$ & $-x_\mu$\\ 
				$\partial^\mu$ & $\partial_\mu$ & $\partial^\mu$ & $-\partial_\mu$  \\
				$\pi^a(x) $	& $\;\;\;-\pi^a(P:x)\;\;\;$ & $\;\;\;-(-1)^a \pi^a(x)\;\;\;$ &  $(-1)^a \pi^a(T:x)$ \\
				$u(x)$	& $u^\dagger(P:x)$ & $u^T(x)$ & $u^*(T:x)$   \\
				$U(x)$	& $U^\dagger(P:x)$ & $U^T(x)$ & $U^*(T:x)$   \\
				$\chi(x)$	& $\chi^\dagger(P:x)$ & $\chi^T(x)$ & $\chi^*(T:x)$   \\
				$s(x)$   & $s^\dagger(P:x)$ & $s^T(x)$ & $s^*(T:x)$  \\ 
				$p(x)$   & $p^\dagger(P:x)$ & $p^T(x)$ & $p^*(T:x)$   \\ 
				$u^\mu(x)$	& $-u_\mu^\dagger(P:x)$ & ${u^\mu}^T(x)$ & $-{u_\mu}^*(T:x)$   \\ 
				$r^\mu(x)$	&$\ell_\mu^\dagger(P:x)$  & $-{\ell^\mu}^T(x)$  & $ r_\mu^*(T:x)$   \\
				$\ell^\mu(x)$	&$r_\mu^\dagger(P:x)$  & $-{r^\mu}^T(x)$  & $ \ell_\mu^*(T:x)$   \\
				
				$v^\mu(x)$	&$v_\mu^\dagger(P:x)$  & $-{v^\mu}^T(x)$  & $ v_\mu^*(T:x)$   \\
				$\;\;\;a^\mu(x)\;\;\;$	&$-a_\mu^\dagger(P:x)$  & ${a^\mu}^T(x)$  & $ a_\mu^*(T:x)$   \\	 
				$\B(x)$	& $\B(P:x)$ & - & $\;\;\;e^{i \pi I^2}e^{i \pi S^2} \B(T:x)\;\;\;$  \\		
				$S^i$		& $S^i$ & - & ${S^i}^*$ \\  
				$I^a$	 		& $I^a$ & - & ${I^a}^*$ \\ 
				$G^{ia}$		&  $G^{ia}$& - &   ${G^{ia}}^*$\\
				\hline \hline
			\end{tabular}
			\caption{P, C and T transformation rules.}
			\label{PCT}
		\end{table}
	\end{center}
\end{widetext}

\section{Composite spin-flavor operators }
\label{app:SFOperators}
An n-body spin-flavor composite operator is defined as the product of n generators of $SU(4)$. In an effective theory such operators will appear with a natural suppression factor $1/N_c^{n-1}$ \cite{Dashen:1993jt,Dashen:1994qi}. For matrix elements in the totally symmetric irrep of $SU(4)$, the 2-body reduction relations in Table \ref{RedRules} are useful in the decomposition of higher body operators into bases operators.
Composite operators can be projected onto definite spin and isospin irreducible tensors, in particular in building a suitable basis. 
Table \ref{SF-basis} gives the basis of composite operators needed in this work. The  notations $X \arrowvert_{J,I}$ indicate projection of the tensor $X$  onto given quantum numbers $(J,I)$.  
The spin-flavor tensors of the UV divergencies of the one-loop diagrams can always be expressed in terms of those bases   operators. This is not the case for the non-analytic terms, which will require explicit evaluations for each of the projections on intermediate baryon states in the loop. For the case of $\pi N\to \pi N$, Appendix \ref{SFreductionspiN} shows those evaluations.
\begin{widetext}
	\begin{center}
		\begin{table}[hbt!]
			\begin{tabular}{c|c}\hline\hline
				\;\; $J\;\; I$\;\; &	Relation  \\ \hline
				0\;\;0 &	$\{S^i,S^i\}-\{I^a,I^a\}=0$ \\
				0\;\;0 &	$\{S^i,S^i\}+\{I^a,I^a\}+4\{G^{ia},G^{ia}\}=\frac{3}{2}N_c(4+N_c)$\\
				0\;\;1 &	$2\{S^i,G^{ia}\}=(2+N_c) I^a $\\
				1\;\;0&	$2\{I^a,G^{ia}\}=(2+N_c)  S^i $\\
				1\;\;1 &	$\frac{1}{2}\{S^k,I^c\}-\eps^{ijk}\eps^{abc}\{G^{ia},G^{jb}\}=(2+N_c)G^{kc}$ \\
				1\;\;1 &	$\eps^{ijk}\{S^i,G^{jc}\}=\eps^{abc}\{I^a,G^{kb}\}$\\
				0\;\;2 &	$4\{G^{ia},G^{ib}\}\arrowvert_{I=2}=\{I^a,I^b\}\arrowvert_{I=2} $\\
				2\;\;0&	$4\{G^{ia},G^{ja}\}\arrowvert_{J=2}=\{S^i,S^j\}\arrowvert_{J=2} $\\ 
				\hline\hline
			\end{tabular}
			\caption{\label{RedRules} $SU(4)$ operator identities    in the totally symmetric irrep $(N_c,0,0)$ of $SU(4)$. The first column gives the composite operator's quantum numbers $(J,I)$ under $SU(2)\times SU(2) $}
			\label{DJM-relations}
		\end{table}
	\end{center}
	\begin{center}
		\begin{table}[h!]
			\begin{tabular}{cc|c|cc|c}
				\hline\hline
				\;J \;  	&  \;I   \;& {\rm Operator} &  	\;J \;  	&  \;I   \;& {\rm Operator} \\
				\hline
				0	&  0& 1 &  2	& 1 &   $\frac {1} {N_c^2} S^iS^j\arrowvert_{J=2} I^a $  \\
				
				1	& 0 & $ S^i $& 2	& 1 & $ \frac {1} {N_c^2} \{S^iS^j\arrowvert_{J=2},G^{ka}\}\arrowvert_{J=2,I=1} $ \\
				
				0	& 1 &$ I^a $ &  1	& 2 & $ \frac {1} {N_c} \{I^a,G^{ib}\}\arrowvert_{I=2}$ \\
				
				1	&  1& $ G^{ia}$&  1	&  2& $ \frac {1} {N_c^2} \{S^i I^a,G^{jb}\}\arrowvert_{J=1,I=2}$  \\
				
				1	&1  & $\frac {1} {N_c} S^i I^a $ & 1	& 2 & $ \frac {1} {N_c^2}S^i I^aI^b\arrowvert_{I=2}$   \\
				
				1	& 1 &  $  \frac {1} {N_c} \{S^i,G^{ja}\}\arrowvert_{J=1}$ & 1	&2  & $\frac {1} {N_c^2} \{I^aI^b\arrowvert_{I=2},G^{ic}\}\arrowvert_{J=1,I=2} $   \\
				
				2	& 0 & $ \frac {1} {N_c} S^i S^j \arrowvert_{J=2} $& 2	&2  & $ \frac {1} {N_c} \{G^{ia},G^{jb}\}\arrowvert_{J=I=2} $ \\
				
				0	& 2 & $\frac {1} {N_c} I^a I^b \arrowvert_{I=2} $  & 	2	& 2 &   $  \frac {1} {N_c^2} \{S^i I^a,G^{jb}\}\arrowvert_{J=I=2}$  \\
				
				2	& 1 & $ \frac {1} {N_c} \{S^i,G^{ja}\}\arrowvert_{J=2}$ & 	2	&2  & \;\;\; $ \frac {1} {N_c^2} \{S^i,  G^{ja}G^{kb}\arrowvert_{J=I=2}\}\arrowvert_{J=I=2}$\;\;\; \\
				
				2	& 1 &  $\frac {1} {N_c^2} \{S^i I^a,G^{jb}\}\arrowvert_{J=2,I=1} $& 2	&  2&  $\frac {1} {N_c^3} S^iS^j\arrowvert_{J=2}I^aI^b\arrowvert_{I=2}$   \\  [.2cm]  
				\hline\hline  \\
			\end{tabular}
			\caption{Spin-flavor tensor operator basis. To complete the basis, to the terms in the table  additional terms involving anti-commutators of those terms with $\frac{1}{N_c^2}\times\hat S^2$ must be added. Most of those terms will be of higher order than the ones needed for renormalizing the scattering amplitudes.}
			\label{SF-basis}
		\end{table}
	\end{center}
\end{widetext}

\section{Lagrangians }
\label{app:Lagrangians}
This Appendix gives the chiral Lagrangians needed in this work, namely the LO Lagrangian for pions, and the LO, NLO and NNLO for baryons, which only include those terms necessary for a complete renormalization of the $\pi B\to \pi B'$ scattering amplitudes. 
The baryon Lagrangians are organized according the $\xi$ power counting. 
\subsection{LO Lagrangians}
The pion LO Lagrangian $\ord{p^2}=\ord{\xi^2}$ has the standard form:
\beq
\mathcal{L}_\pi^{(2)}=\frac{1}{4}F_\pi^2
\langle D_\mu U^\dag D^\mu U+\chi U^\dag+\chi^\dag U\rangle.
\eeq

The   LO baryon Lagrangian is $\ord{\xi}$ and given by:
\begin{widetext}
\beq
\mathcal{L}_B^{(1)}=\Bdag\Big(i D_0+\gA u^{i a} G^{i a}-\frac{C_{H F}}{N_c}  \hat S^2+\frac{c_1 N_c}{2 \Lambda}\langle{\chi}_{+}\rangle\Big) \B.
\eeq

\begin{center}
\begin{table}[t]
	\begin{tabular}{lll}
\vspace*{-2cm}			
		\btik
		\begin{feynman}	
			\vertex (a0) ;	
			\vertex[above left=1.25cm of    a0] (a1){\(k_1  a\)};
			\vertex[above right=1.25cm of    a0] (a2){\(k_2  b\)};
			\vertex[below left=1.25cm of    a0] (a3){\(k_3  c\)};
			\vertex[below right=1.25cm of    a0] (a4){\(k_4  d\)};
			\diagram* {{(a0) -- [charged scalar,dashed, thick] (a1),
					(a0) -- [charged scalar,dashed, thick] (a2),
					(a0) -- [charged scalar,dashed, thick] (a3),
					(a0) -- [charged scalar,dashed, thick] (a4) }};
		\end{feynman}
		\etik & 
	\(	\begin{array}{l} =\frac{i}{3F_\pi^2}\Big(M_\pi^2(\delta^{ab}\delta^{cd}+\delta^{ac}\delta^{bd}+\delta^{ad}\delta^{bc})\\
			\;\;\;\;	+\delta^{ab}\delta^{cd}(-(k_1+k_2)\cdot(k_3+k_4)+2(k_1\cdot k_2+k_3\cdot k_4)) \\ \;\;\;\; +\delta^{ac}\delta^{bd}(-(k_1+k_3)\cdot(k_2+k_4)+2(k_1\cdot k_3+k_2\cdot k_4))\\ \;\;\;\; +\delta^{ad}\delta^{bc}(-(k_2+k_3)\cdot(k_1+k_4)+2(k_1\cdot k_4+k_2\cdot k_3))\Big)\\ \\ \\ \\ \\  \end{array} \)&\\
		\btik
		\begin{feynman}
			\vertex (a1) {\(p\)};
			\vertex[right=1.5cm of  a1] (a2);
			\vertex[right=1.5cm of a2] (a3);
			\vertex[below=1.5cm of a2](b1);
			\diagram* {{[edges=fermion]
					(a1) -- (a2) -- (a3)  },
				(a2) -- [charged scalar,dashed, thick,edge label=\(k  a\)] (b1)};
		\end{feynman}
		\etik
		& 
	\(	\begin{array}{l} =\frac{\gA}{F_\pi} k^i G^{ia} \\ \\ \\   \end{array}\)
		&
		\\
		
		\btik
		\begin{feynman}
			\vertex (a1) ;
			\vertex[right=1.5cm of a1] (a2);
			\vertex[right=1.5cm of a2] (a3);
			\vertex[below left=1.5cm of a2](b1){\(k_1 a\)};
		\vertex[below right=1.5cm of a2] (b2){\(k_2 b\)};
	\diagram* {{[edges=fermion]
			(a1) -- (a2) -- (a3) },
		(a2) -- [charged scalar,dashed, thick] (b1),(a2) -- [charged scalar,dashed, thick] (b2)};
\end{feynman}
\etik 
& 
\(\begin{array}{l}  =\frac{1}{F_\pi^2}\Big( \frac 12 (k_2^0-k_1^0)\eps^{abc} I^c-i 2  c_1 N_c \frac{M_\pi^2}{\Lambda} \delta^{ab}\Big)  \\ \\ \\ \end{array}\)&\\
\vspace*{-1cm}			
\btik
\begin{feynman}
\vertex (a1) ;
\vertex[right=1.5cm of a1] (a2);
\vertex[right=1.5cm of a2] (a3);
\vertex[below left=1.5cm of a2](b1){\(k_1 a\)};
\vertex[below right=1.5cm of a2] (b2){\(k_2 b\)};
\vertex[above =1.5cm of a2] (b3){\(k_3 c\)};
\diagram* {{[edges=fermion]
(a1) -- (a2) -- (a3) },
(a2) -- [charged scalar,dashed, thick] (b1),(a2) -- [charged scalar,dashed, thick] (b2),(a2) -- [charged scalar,dashed, thick] (b3)};
\end{feynman}
\etik
& 
\(\begin{array}{l}  =-\frac{\gA}{F_\pi^3}G^{id}\Big(   \delta^{ab}\delta^{cd} (2 k_3^i-k_1^i-k_2^i)\\
\;\;\;\; +\delta^{ac}\delta^{bd} (2 k_2^i-k_1^i-k_3^i)
+\delta^{ad}\delta^{bc} (2 k_1^i-k_2^i-k_3^i)\Big)  \\ \\ \\ \\ \end{array}\)&\\
\vspace*{-2cm}
\btik
\begin{feynman}
\vertex (a1) ;
\vertex[right=1.5cm of a1] (a2);
\vertex[right=1.5cm of a2] (a3);
\vertex[below left=1.5cm of a2](b1){\(k_3 c\)};
\vertex[below right=1.5cm of a2] (b2){\(k_4 d\)};
\vertex[above left =1.5cm of a2] (b3){\(k_1 a\)};
\vertex[above right =1.5cm of a2] (b4){\(k_2 b\)};
\diagram* {{[edges=fermion]
(a1) -- (a2) -- (a3) },
(a2) -- [charged scalar,dashed, thick] (b1),(a2) -- [charged scalar,dashed, thick] (b2),(a2) -- [charged scalar,dashed, thick] (b3),(a2) -- [charged scalar,dashed, thick] (b4)};
\end{feynman}
\etik
& 
\(\begin{array}{l}  = \frac{1}{F_\pi^4}\Big(\frac{1}{12 }I^{e}\Big(   \delta^{ab}\eps^{cde} (k_3^0-k_4^0)+
\delta^{ac}\eps^{bde} (k_2^0-k_4^0)+
\delta^{ad}\eps^{bce} (k_2^0-k_3^0)\\
\;\;\;\;\; +
\delta^{bc}\eps^{ade} (k_1^0-k_4^0)+
\delta^{bd}\eps^{ace} (k_1^0-k_3^0)+
\delta^{cd}\eps^{abe} (k_1^0-k_2^0)\Big)  \\ 
\;\;\;\;\;+i\frac 23 c_1 N_c \frac{M_\pi^2}{\Lambda}(\delta^{ab}\delta^{cd}+\delta^{ac}\delta^{bd}+\delta^{ad}\delta^{bc})\Big)\\ \\ \\ \\
\end{array}\)&
\end{tabular}
\caption{Vertices from the LO Lagrangians.}
\label{vertices}
\end{table}
\end{center}
\end{widetext}
The  vertices from the LO Lagrangian needed in the NNLO   calculations are depicted in  Table \ref{vertices}.

\subsection{Higher order Lagrangians}

In the construction of the higher order Lagrangians one uses the LO equations of motion, namely:
\bea
iD_0 \B&=&\left(\frac{C_{\rm HF}}{N_c} \hat S^2-c_1 \frac{N_c}{2\Lambda}\langle \chi_+\rangle-\gA u^{ia} G^{ia}\right)\B\nonumber\\
D_\mu u^\mu&=& \frac i 2 \chi_-,
\eea
and the identities:
\bea
D_\mu u_\nu-D_\nu u_\mu &=&-f_{-\mu\nu}\nonumber\\
\; [D_\mu,D_\nu]&=& -i \Gamma_{\mu\nu}\nonumber\\
\Gamma_{\mu\nu}&=& \frac 12 f_{+\mu\nu}+\frac i 4 [u_\mu,u_\nu].
\eea
The following are the higher order Lagrangians needed for renormalization in the present calculation of the  $\pi B\to \pi B'$ amplitudes. 
 The CT Lagrangians needed to renormalize UV divergencies are organized by the order in $\gA$ of the UV divergencies, and are the following:
 \begin{widetext}
\bea
\mathcal{L}_{C T}\left(\gA^4\right)&=&\Bdag\left(\frac{f_{10}}{\Lambda^2} \eps^{i j k} S^k u^{ia} \overleftrightarrow{D}_0 u^{ja}\right.  +\frac{f_{01}}{\Lambda^2} \eps^{a b c} I^c u^{ia} \overleftrightarrow{D}_0 u^{ib} \nonumber\\
&+&\frac{f_{11}^{(1)}}{\Lambda} \eps^{i j k} \eps^{a b c} u^{ia} u^{jb} G^{k c}  +\frac{f_{11}^{(2)}}{\Lambda} \eps^{i j k} \eps^{a b c} u^{ia} u^{jb} \frac{1}{N_c}S^k I^c \nonumber \\
& +&\left.\frac{f_{20}}{\Lambda} u^{ia} u^{ja} \frac{1}{N_c} S^i S^j\right|_{J=2}   +\left.\frac{f_{02}}{\Lambda} u^{ia} u^{ib} \frac{1}{N_c} I^a I^b\right|_{I=2}\nonumber \\
& +& \left.\left.\frac{f_{22}}{\Lambda} u^{ia} u^{jb} \frac{1}{N_c} G^{i a} G^{j b}\right|_{J=I=2}\right) \B
\label{UV-CTs4}
\eea
\bea
\mathcal{L}_{C T}\left(\gA^2\right)&=&\Bdag\left(g^{(1)}_{00}  \frac{\left\langle \chi_{+}\right\rangle}{\Lambda}+g_{00}^{(2)}   \frac{\left\langle \chi_{+}\right\rangle}{\Lambda} \frac{\hat S^2}{N_c}\right. \nonumber\\
& +&\frac{g_{00}^{(3)}}{{\Lambda}} u_\mu^a u^{a \mu}+\frac{g_{00}^{(4)}}{\Lambda} u_\mu^a u^{a \mu} \frac{ \hat S^2}{N_c} \nonumber\\
& +& i \frac{g_{01}^{(1)}}{\Lambda^2} \eps^{a b c} \chi_{-}^a u_0^b I^c+\frac{g_{01}^{(2)}}{\Lambda^2} \eps^{a b c} u_\mu^a \overleftrightarrow{D}_0 u^{b \mu} I^c \nonumber\\
& +&\left.\frac{g_{0 2}^{(1)}}{\Lambda} u_\mu^a u^{b \mu} I^a I^b\right|_{I=2} \nonumber\\
& +&\frac{g_{11}^{(1)}}{\Lambda} \eps^{i j k} \eps^{a b c} u^{i a} u^{j b} G^{k c}   \left.+\frac{g_{11}^{(2)}}{\Lambda} \eps^{i j k} \eps^{a b c} u^{i a} u^{j b} \frac{1}{N_c} S^k I^c\right) \B 
\label{UV-CTs2}
\eea
\bea
{\mathcal{L}}_{CT}\left(\gA^0\right)&=&\Bdag\left({h}_{00}^{(1)} \frac{N_c}{\Lambda^3}\left\langle \chi_{+}^2\right\rangle+h_{00}^{(2)} \frac{N_c}{\Lambda^3}\left\langle\chi_{-}^2\right\rangle\right. +h_{00}^{(3)} \frac{N_c}{\Lambda^3}\left\langle\chi_{+}\right\rangle u_\mu^a u^{\mu a}\nonumber\\
&+&\left. i h_{01}^{(1)} \frac{\eps^{a b c}}{\Lambda^2} \chi_{-}^a u_0^b I^c +h_{01}^{(2)} \frac{\eps^{a b c}}{\Lambda^2} u_\mu^a \overleftrightarrow{D}_0 u^{\mu b} I^c\right.\nonumber\\
&+&\left. h_{01}^{(3)} \frac{\eps^{a b c}}{\Lambda^2} u_0^a \overleftrightarrow{D}_0 u_0^{  b} I^c \right)\B.
\label{UV-CTs0}
\eea
\end{widetext}
The LEC notation explicitly gives the $(J,I)$ of the spin-flavor tensor, namely $LEC_{JI}$. Note that a few terms have the same structure across the depicted Lagrangians, thus the corresponding LECs add up. They have been presented separately because in the renormalization the corresponding $\beta$ functions are organized in powers of $\gA$.

In addition, and to the order of the present calculations, there are  Lagrangian terms that serve as finite CTs. Here again only those contributing to the $\pi B\to \pi B'$ amplitudes are shown.
\begin{widetext}
\bea
\mathcal{L}_{C T}(\text{finite})&=& 
i\frac{\ell_{00}^{(1)}}{\Lambda^2} u_0^a u^{ia} \;\;\Bdag \overleftrightarrow{D}^i\B
+i\frac{\ell_{00}^{(2)}}{\Lambda^2} \langle \chi_+\rangle  \Bdag \overleftrightarrow{D}_0\B \nonumber\\
&+&\ell_{01}^{(1)} \frac{1}{2 m_0} \Bdag \vec D^2 \B
+      \frac{\ell_{01}^{(2)}}{\Lambda^2} \eps^{abc} u^{ia} D^i u_0^b  \;\;\Bdag I^c\B 
+\frac{\ell_{01}^{(3)} }{\Lambda^2}
\eps^{abc}  D^i u^{ia}  u_0^b  \;\;\Bdag I^c   \B
\nonumber\\
&+&\frac{\ell_{10}^{(1)}}{\Lambda^2} \eps^{ijk}u^{ia}{D}^j u_0^a  \;\;\Bdag S^k\B
+\frac{\ell_{10}^{(2)}}{\Lambda^2} \eps^{ijk} {D}^j u^{ia} u_0^a  \;\;\Bdag S^k\B
\nonumber\\
&+&i\frac{\ell_{11}^{(1)}}{2\Lambda^2}\eps^{ijk}\eps^{abc} u_0^a u^{ib}  \;\;\Bdag  \{\overleftrightarrow{D}^j, 
G^{kc}\}\B
+i\frac{\ell_{11}^{(2)}}{2\Lambda^2} \eps^{ijk}\eps^{abc}  u^{ia}  u^{jb}  \;\;\Bdag \{\overleftrightarrow{D}_0, G^{kc}\}\B\nonumber\\
&+& \text{ terms with either $J$ or $I$ bigger than 1},
\label{Finite-CTs}
\eea
\end{widetext}
where the terms   not contributing to the $\pi N\to \pi N$ amplitude are not explicitly shown, i.e. terms where $J$ or $I$ are larger than 1.

For the  contributions to the scattering amplitudes, the renormalization of the baryon masses and the $\pi B$ couplings are needed. The corresponding CT Lagrangians are the following: 
\begin{widetext}

\bea
{\cal{L}}{  _{m}^{CT}}&=&\Bdag \Big(  \frac{C_{H F 1}}{N_c^2}  \hat S^2+\frac{C_{H F 2}}{N_c^3}  \hat S^4+ \frac{c_2}{N_c} \langle  \chi_{+} \rangle   \hat S^2
 \Big)\B.
 \eea


\bea
{\cal{L}}{  _{A}^{CT}}&=&\Bdag\Big( u_{ia}\left( \frac{C^A_0}{N_c} G^{ia}+\frac{C^A_1}{4\Lambda^2}\{\chi_+,G^{ia}\}+ \frac{C^A_2}{N_c^2} \{ \hat S^2,G^{ia}\}+ \frac{C^A_3}{N_c} [ \hat S^2,G^{ia}]+ 
\frac{C^A_4}{N_c} S^iI^a\right) \nonumber\\
&+&  i\; \frac{C^A_5}{2\Lambda^2}D^i \chi_-^a G^{ia}
\Big)\B+
 i\; \frac{\gA}{m_0} u^{0a}\Bdag  \{G^{ia}, \overleftrightarrow{D}^i\} \B.
\label{eq:axial-current-CT-lagrangian}
\eea
\end{widetext}
This Lagrangian provides the renormalization CTs for both the axial current and the $\pi B$ interaction. The loop correction preserves the Goldberger-Treiman relation, whose discrepancy is only due to the  term proportional to $C^A_5$. The term proportional to  $\gA/m_0$, demanded by the non-relativistic expansion, gives the leading contribution to the time component of the axial current and the corresponding pion-baryon coupling, giving contributions $\ord{\xi^2}$ with respect to the spatial components of the axial current. Although these contributions to the pion-baryon coupling are NNLO, they are found   to be virtually insignificant and are thus disregarded in the fits to the data.

\section{Loop integrals }
\label{app:Integrals}
This Appendix gives details on the loop integrals, where throughout dimensional regularization is used, with $d=4-2\eps$, and the notation $\lambda_\eps\equiv  \frac{1}{\eps} -\gamma_E+\log 4\pi$.

All one-loop integrals, after convenient integration variable shifts, can be brought to the following general well known forms, leaving integrations over Feynman parameters to be performed:
\beq
\int \widetilde{d^d k}\frac{\{1; k^\mu k^\nu;\cdots\}}{(k^2-\Lambda^2+i\eps)^n},
\eeq
 where $\widetilde{d^d k}=\frac{d^d k}{(2\pi)^d}$, and $\Lambda$ contains in particular dendencies on Feynman parameters.
 
Diagrams not involving heavy baryon propagators lead straightforwardly to the integral forms shown above. When heavy baryon propagators are involved, the general form of the loop integrals are obtained by first using the Feynman parameter representation:
\begin{widetext}
\bea
	  \frac{1}{A_1 \cdots A_m a_1 \cdots a_n}&=& 
  2^m \Gamma(m+n)  \int_0^1 d \alpha_1 \int_0^{1 } d \alpha_2 \cdots \int_0^{1 } d \alpha_{n } \\
	&\times & \int_0^{\infty} d \lambda_1 \cdots \int_0^{\infty} d \lambda_m\;\frac{\delta(1-\alpha_1-\cdots-\alpha_n)}{\left(2 \lambda_1  A_1+\cdots+\alpha_1 a_1+\cdots  \right)^{(n+m)}}\nonumber
\eea
\end{widetext}
The cases needed in this work are those with one, two and three heavy baryon propagators. All cases can be brought to the case with a single heavy baryon propagator using the partial fraction decomposition. Thus the integrals of interest will be of the general form:
\begin{widetext}
\bea
&&\int \widetilde{d^d k} \frac{{\rm Num}(k,\{q_i\})}{(Q^0-k^0+i\eps)(k^2-M_\pi^2+i\eps)((k-q_1)^2-M_\pi^2+i\eps)\cdots((k-q_n)^2-M_\pi^2+i\eps)}\\
&=&2 \Gamma(n) \int_0^\infty d\lambda \int_0^1 d\alpha_1 \cdots d\alpha_n \int \widetilde{d^d k} \frac{\delta(1-\alpha_1-\cdots-\alpha_n){\rm Num}(k,\{q_i\})}{(2\lambda (Q^0-k^0+i\eps)+\sum_i^n \alpha_i ((k-q_i)^2-M_\pi^2+i\eps))^{n+1}},\nonumber
\eea
\end{widetext}
where ${\rm Num}$ is a polynomial of the arguments.
Upon integrating over $k$, the integration over the Feynman parameter $\lambda$ has  the general form:
\beq
J(r,\nu,C_0,C_1,\lambda_0)\equiv\int_0^\infty \!\!\! d\lambda \frac{\lambda^r}{(C_0+C_1\;(\lambda-\lambda_0)^2)^\nu},
\eeq
where $r=0,1,\cdots$, and $\nu={\rm integer}-\frac d 2$.

\begin{widetext}
The integrals $J(r,\nu,C_0,C_1,\lambda_0)$ satisfy the following recurrence relations:

\bea
J(0,\nu,C_0,C_1,\lambda_0)&=& \frac{1}{2\nu-1}(2\nu C_0 J(0,\nu+1,C_0,C_1,\lambda_0)-\frac{\lambda_0}{(C_0+C_1 \lambda_0^2)^\nu})\nonumber\\
J(1,\nu,C_0,C_1,\lambda_0)&=&\lambda_0 J(0,\nu,C_0,C_1,\lambda_0)+  \frac{(C_0+C_1\lambda_0^2)^{1-\nu}}{2C_1(\nu-1)}\nonumber\\
J(2,\nu,C_0,C_1,\lambda_0)&=&(\lambda_0^2-C_0) J(0,\nu,C_0,C_1,\lambda_0)\nonumber\\
&+&J(0,\nu-1,C_0,C_1,\lambda_0)+\lambda_0 \frac{(C_0+C_1\lambda_0^2)^{1-\nu}}{2C_1(\nu-1)}
\label{Jintegrals1}
\eea
The seed integral is taken to be the one for $\nu=1+\eps$:
\bea
J(0,1+\eps,C_0,C_1,\lambda_0)&=&\frac {1}{\sqrt{C_0C_1}}\Big( \frac\pi 2 + \arctan\frac{\lambda_0}{ \sqrt{C_0/C_1}}\Big)\nonumber\\
 &-&\eps
 \int_0^\infty \frac{\log(C_0+C_1(\lambda-\lambda_0)^2)}{C_0+C_1(\lambda-\lambda_0)^2} d\lambda
 \label{Jintegrals2}
 \eea
 \end{widetext}
 The integrals $J(0,n+\eps,C_0,C_1,\lambda_0)$ with $n=0,-1,-2,\cdots$ are polynomials in $C_0$, $C_1$ and $\lambda_0$, while for $n>0$ they do contain   non-analytic terms. In general, the integrals always appear in the one-loop results in the combination: $\Gamma(\nu)J(n,\nu,C_0,C_1,\lambda_0)$. 
 In the actual calculations of this work, it turns out that $C_1=1$ throughout.
 
With these results,  Table \ref{lambda-integrals} yields the necessary $\lambda$-Feynman parameter integrals for the present one-loop calculations.

\begin{widetext}
\begin{center}
\begin{table}[t]
\renewcommand*{\arraystretch}{1.5}
\begin{tabular}{cc|c}
	\hline\hline
$\;\;\;	r\;\;\;$ & $\;\;\; \nu\;\;\;$ & $ \Gamma(\nu+\eps)J(r,\nu+\eps,C_0,C_1,\lambda_0)$ 
\\ \hline  
	0 & 0 & $ \;\;\;\frac{1}{2} \left(\lambda_0^2-\frac{C_0}{C_1}\right) \lambda_\eps -\frac{\sqrt{C_0 C_1}
		\lambda_0 \left(2 \arctan\left(\sqrt{\frac{C_1}{C_0}} \lambda_0\right)+\pi \right)}{C_1}+\frac{3
		C_1 \lambda_0^2-C_0+\left(C_0-C_1 \lambda_0^2\right) \log \left(C_1 \lambda_0^2+C_0\right)}{2 C_1} $  \\
	0 & 1 & $ \frac{\lambda_\eps }{2 C_1}+\frac{\lambda_0 \left(2 \arctan\left(\sqrt{\frac{C_1}{C_0}}
		\lambda_0\right)+\pi \right)}{2 \sqrt{C_0 C_1}}-\frac{\log \left(C_1 \lambda_0^2+C_0\right)}{2C_1} $ \\
	0 & 2 & $ \frac{\lambda_0 \left(2 \arctan\left(\sqrt{\frac{C_1}{C_0}} \lambda_0\right)+\pi \right)}{4	C_0 \sqrt{C_0 C_1}}+\frac{1}{2 C_0 C_1} $ \\
	0 & 3 & $ \frac{3 C_1 \lambda_0^2+2 C_0}{4 C_0^2 C_1 \left(C_1 \lambda_0^2+C_0\right)}+\frac{3 \lambda_0 \left(2 \arctan\left(\sqrt{\frac{C_1}{C_0}} \lambda_0\right)+\pi \right)}{8 C_0^2 \sqrt{C_0 C_1}} $ \\
	1 & 0 & $\;\;\; \frac{1}{2} \left(\lambda_0^2-\frac{C_0}{C_1}\right) \lambda_\eps -\frac{\sqrt{C_0 C_1}
		\lambda_0 \left(2 \arctan\left(\sqrt{\frac{C_1}{C_0}} \lambda_0\right)+\pi \right)}{C_1}+\frac{3
		C_1 \lambda_0^2-C_0+\left(C_0-C_1 \lambda_0^2\right) \log \left(C_1 \lambda_0^2+C_0\right)}{2 C_1} $ \\
	1 & 1 & $ \frac{\lambda_\eps }{2 C_1}+\frac{\lambda_0 \left(2 \arctan\left(\sqrt{\frac{C_1}{C_0}}
		\lambda_0\right)+\pi \right)}{2 \sqrt{C_0 C_1}}-\frac{\log \left(C_1 \lambda_0^2+C_0\right)}{2
		C_1} $ \\
	1 & 2 & $ \frac{\lambda_0 \left(2 \arctan\left(\sqrt{\frac{C_1}{C_0}} \lambda_0\right)+\pi \right)}{4
		C_0 \sqrt{C_0 C_1}}+\frac{1}{2 C_0 C_1} $ \\
	1 & 3 & $ \frac{3 C_1 \lambda_0^2+2 C_0}{4 C_0^2 C_1 \left(C_1 \lambda_0^2+C_0\right)}+\frac{3 \lambda_0 \left(2 \arctan\left(\sqrt{\frac{C_1}{C_0}} \lambda_0\right)+\pi \right)}{8 C_0^2 \sqrt{C_0 C_1}} $ \\
	2 & 0 & $\;\;\; \frac{1}{2} \left(\lambda_0^2-\frac{C_0}{C_1}\right) \lambda_\eps -\frac{\sqrt{C_0 C_1}
		\lambda_0 \left(2 \arctan\left(\sqrt{\frac{C_1}{C_0}} \lambda_0\right)+\pi \right)}{C_1}+\frac{3
		C_1 \lambda_0^2-C_0+\left(C_0-C_1 \lambda_0^2\right) \log \left(C_1 \lambda_0^2+C_0\right)}{2 C_1} $ \\
	2 & 1 & $ \frac{\lambda_\eps }{2 C_1}+\frac{\lambda_0 \left(2 \arctan\left(\sqrt{\frac{C_1}{C_0}}
		\lambda_0\right)+\pi \right)}{2 \sqrt{C_0 C_1}}-\frac{\log \left(C_1 \lambda_0^2+C_0\right)}{2
		C_1} $ \\
	2 & 2 & $ \frac{\lambda_0 \left(2 \arctan\left(\sqrt{\frac{C_1}{C_0}} \lambda_0\right)+\pi \right)}{4
		C_0 \sqrt{C_0 C_1}}+\frac{1}{2 C_0 C_1} $ \\
	2 & 3 & $ \frac{3 C_1 \lambda_0^2+2 C_0}{4 C_0^2 C_1 \left(C_1  \lambda_0^2+C_0\right)}+\frac{3 \lambda_0 \left(2 \arctan\left(\sqrt{\frac{C_1}{C_0}} \lambda_0\right)+\pi \right)}{8 C_0^2 \sqrt{C_0 C_1}} $ \\ \hline\hline
\end{tabular}
\caption{Integrals over the Feynman parameter $\lambda$.}
\label{lambda-integrals}
\end{table} 
\end{center}

A particular integral, which is convenient to display explicitly is:
\bea
\cal{I}(Q,M_\pi)&=&\frac{1}{d-1}\int \widetilde{d^d k}\;\frac{\vec k^2}{(k^0-Q+i\eps)(k^2-M_\pi^2+i\eps)}\nonumber\\
&=&
\frac{i}{3(4\pi)^2}\bigg(Q\left((3M_\pi^2-2Q^2)\Big(\lambda_\eps-\log \frac{M_\pi^2}{\mu^2}\Big)+7M_\pi^2-\frac{16}{3}Q^2\right) \nonumber\\
&+&  4(M_\pi^2-Q^2-i\eps)^{3/2}\Big(\frac \pi 2- \arctan\frac{Q}{\sqrt{M_\pi^2-Q^2-i\eps}}\Big)\bigg)
\eea
\end{widetext}
This integral appears in the self-energy and in several of the one-loop contributions to the scattering amplitude.

\section{NNLO Scattering Amplitudes}\label{app:NNLO-amplitudes}

This Appendix gives details of the calculations of the one-loop diagrams for the scattering amplitudes, organized by their power in $\gA$. The UV divergent pieces are analyzed in detail for the purpose of demonstrating the large $N_c$ consistency of the results. Also included are the higher order Lagrangian contributions to the amplitudes, as well as the results for the $\beta$ functions of the LECs. The  simplifications for the case of $\pi N\to \pi N$ scattering are also included, along with results for the reductions of composite spin-flavor tensors needed in that case.

The case of $\pi N \to \pi N$ scattering shows significant simplifications, shown explicitly for the CT contributions.  The reductions of composite spin-flavor tensors needed in that case are also included in subsection \ref{SFreductionspiN}.

\subsection{Diagrams $ \boldsymbol{\propto \mathring g_A^4}$}

The diagrams proportional to $\gA^4$ are shown in Fig.(\ref{Fig-loops-gA4}).
Diagrams $D_1$, $D_2$  and $D_3$ will contribute with pole terms, i.e., terms that contain singularities due to a single baryon pole. Diagram $D_1$ has   double pole,   single pole,  and   no-pole contributions, while diagrams $D_2$ and $D_3$ have single pole and no-pole contributions.

First consider diagrams $D_2$: they have the general structure of a diagram shown in   Fig.(\ref{polediagrams} (a)). The amplitude corresponding to that diagram reads:
\begin{widetext}
\bea
\label{apole}
i T(a)&=&-\frac{i}{p_1^0+k_1^0-\delta m_n}\Gamma_2(p_1^0+k_1^0,p_2^0,k_2){\cal{P}}_n \Gamma_1(p_1^0,p_1^0+k_1^0,-k_1)
\\
&=&-i\left(\frac{\Gamma_2(\delta m_n,p_2^0,k_2){\cal{P}}_n \Gamma_1(p_1^0,\delta m_n,-k_1)}{p_1^0+k_1^0-\delta m_n}\right.\nonumber \\
&+&\left.\frac{\Gamma_2(p_1^0+k_1^0,p_2^0,k_2){\cal{P}}_n \Gamma_1(p_1^0,p_1^0+k_1^0,-k_1)-\Gamma_2(\delta m_n,p_2^0,k_2){\cal{P}}_n \Gamma_1(p_1^0,\delta m_n,-k_1) }{p_1^0+k_1^0-\delta m_n} \right), 
\nonumber
\eea
\end{widetext}
where in the last expression the first term contains a pole and the last one does not. When renormalizing, the first term will only require the renormalization of the vertices, while the last one will require renormalization of the scattering amplitudes, i.e., local terms with two pions.

\begin{widetext}
	\begin{center}
		\begin{figure}[b]
			\centerline{\includegraphics[width=12.cm,angle=-0]{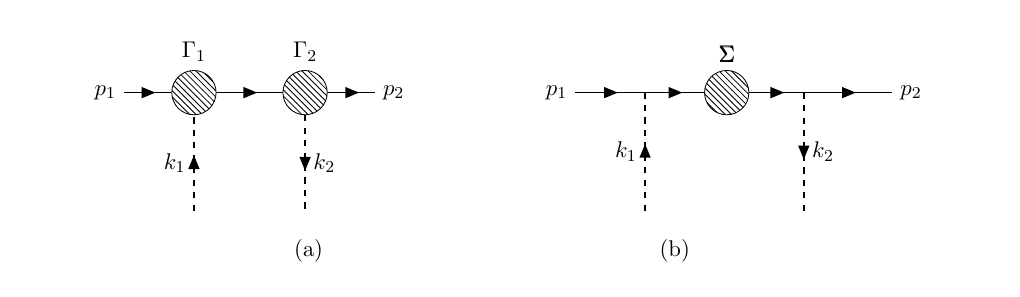}}
			\caption{Pole diagrams.}
			\label{polediagrams}
		\end{figure}
	\end{center}
\end{widetext}

The case of the diagram Fig.(\ref{polediagrams} (b)) with double pole is decomposed into terms with   double pole,  single pole and  no-pole, namely:
\begin{widetext}
\bea
i T(b)&=&  \Gamma_2 {\cal{P}}_n \Sigma(p_1^0+k_1^0){\cal{P}}_n \Gamma_1 \frac{1}{(p_1^0+k_1^0-\delta m _n)^2}\nonumber\\
&=& \frac{1}{(p_1^0+k_1^0-\delta m_n)^2}\Gamma_2 {\cal{P}}_n\left(  \Sigma(\delta m_n)+(p_1^0+k_1^0-\delta m_n)  \Sigma'(\delta m_n)\right.\nonumber\\
&+& \left.(\Sigma(p_1^0+k_1^0)-\Sigma(\delta m_n)-(p_1^0+k_1^0-\delta m_n)  \Sigma'(\delta m_n))\right)
{\cal{P}}_n\Gamma_1,
\label{bpole}
\eea
\end{widetext}
where one identifies: $\Sigma'(\delta m_n)=-\delta Z(\delta m_n)$. The first term has a double pole which is taken care of by the mass renormalization, the second term has a single pole, and the last one has no-pole. For the single pole term the renormalization is through the vertex renormalization.

Applying those decompositions to the corresponding diagrams leads to the following results:

\begin{widetext}
{Diagram $D_1$:}

\bea
i T_{D_1}^{ba}&=&-i \left(\frac{\mathring g_A}{F_\pi} \right)^2 k_1^i k_2^j \sum_{n}\;G^{jb}{\cal{P}}_nG^{ia}  \frac {  \Sigma_n(p_1^0+k_1^0)}{(p_1^0+k_1^0-\delta m_n)^2}\nonumber\\
&=& -i \left(\frac{\mathring g_A}{F_\pi} \right)^2 k_1^i k_2^j\; \sum_{n}G^{jb}{\cal{P}}_nG^{ia} \left(\frac { \Sigma_n(\delta m_n)  }{(p_1^0+k_1^0-\delta m_n)^2}-\frac {\delta Z_n(\delta m_n) }{p_1^0+k_1^0-\delta m_n}\right.\nonumber\\
&+&\left.\frac {\Sigma_n(p_1^0+k_1^0) - \Sigma_n(\delta m_n)}{(p_1^0+k_1^0-\delta m_n)^2}+\frac {\delta Z_n(\delta m_n) }{p_1^0+k_1^0-\delta m_n}\right)
\eea

{Diagrams $D_2$:}

\bea
i T_{D_2}^{ba}&=&\frac{\mathring g_A}{F_\pi} k_1^i k_2^j\sum_{n} \frac{1}{p_1^0+k_1^0-\delta m_n}
\left( G^{jb} \proj{n} \Gamma^{ia}(p_1^0,p_1^0+k_1^0,-k_1)+\Gamma^{jb}(p_1^0+k_1^0,p_2^0, k_2)\proj{n}G^{ia}\right)\nonumber\\
&=& \frac{\mathring g_A}{F_\pi} k_1^i k_2^j \sum_{n}\frac{1}{p_1^0+k_1^0-\delta m_n}
\left(  G^{jb} \proj{n} \Gamma^{ia}(p_1^0,\delta m_n,-k_1)+\Gamma^{jb}(\delta m_n,p_2^0, k_2)\proj{n}G^{ia}\right.\\
&+&   \left(  G^{jb} \proj{n}( \Gamma^{ia}(p_1^0,p_1^0+k_1^0,-k_1)-\Gamma^{ia}(p_1^0,\delta m_n,-k_1))\right.\nonumber\\
&+&\left. \left. (\Gamma^{jb}(p_1^0+k_1^0,p_2^0, k_2)-\Gamma^{jb}(\delta m_n,p_2^0, k_2))\proj{n}G^{ia}\right)\right)\nonumber
\eea

{Diagrams $D_3$:}

\bea
iT_{D_3}^{ba}&=& -i \left( \frac{\mathring g_A}{F_\pi} \right)^2 k_1^i k_2^j \sum_{n}\;G^{jb} \proj{n}G^{ia}\frac{1}{p_1^0+k_1^0-\delta m_n}\left( \frac{\Sigma_{in}(\delta m_{in})}{p_1^0-\delta m_{in}}+\frac{(\Sigma_{out}(\delta m_{out})}{p_2^0-\delta m_{out}}\right.\nonumber\\
&+&\left. \frac{ \Sigma_{in}(p_1^0)-\Sigma_{in}(\delta m_{in}}{p_1^0-\delta m_{in}}+\frac{\Sigma_{out}(p_2^0)-\Sigma_{out}(\delta m_{out})}{p_2^0-\delta m_{out}}\right)
\eea
\end{widetext}
The term with external pole will be absorbed by renormalizing the baryon masses of the external baryons, and the terms with no external pole multiplied by 1/2, which are a consequence of the external baryon's wave function renormalization, are part of the scattering amplitude, which  reduces to:
\begin{widetext}
\beq
i T_{D_3}^{ba}= \frac i 2  \left( \frac{\mathring g_A}{F_\pi} \right)^2 k_1^i k_2^j\sum_{n} \frac{1}{p_1^0+k_1^0-\delta m_n} \{G^{jb}\proj{n}G^{ia},\delta \hat Z\}
\eeq

{Diagram $D_4$:} 

\bea
T_{D_4}^{ba}&=& \left( \frac{\mathring g_A}{F_\pi} \right)^4k_1^i k_2^j\sum_{n,n',n''}
\;G^{lc}\proj{n''}G^{jb}\proj{n'} G^{ia} \proj{n} G^{lc}\nonumber\\
&\times&  \left(\frac{{\cal{I}}(\delta m_n-p_1^0 ,M_\pi)}{(p_2^0-p_1^0-\delta m_{n''}+\delta m_n)(k_1^0+\delta m_n-\delta m_{n'})}\right.\\
&+&\left.\frac{{\cal{I}}( \delta m_{n''}-p_2^0,M_\pi)}{(p_2^0-p_1^0-\delta m_{n''}+\delta m_n)(p_2^0-p_1^0-k_1^0-\delta m_{n''}+\delta m_{n'})}\right.\nonumber\\
&-&\left.\frac{{\cal{I}}(\delta m_{n'}-p_1^0-k_1^0 ,M_\pi)}{(p_2^0-p_1^0-k_1^0-\delta m_{n''}+\delta m_{n'})(k_1^0+\delta m_n-\delta m_{n'})}\right)\nonumber
\eea
\end{widetext}

\subsubsection{\rm UV divergencies of the no-pole terms}
The UV divergent terms of the no-pole terms of the diagrams proportional to $ \mathring g_A^4$ are as follows:

\begin{widetext}
{Diagram $D_1$ UV:}

\bea
i T_{(D_1)}^{ba^{UV}}({\rm no-pole})&=&i\lepspi \left(\gAoF \right)^4\frac 23 k_1^i k_2^j\sum_{n,n'} (\frac 12(k_1^0+k_2^0)+\frac 12(\delta m_{in}+\delta m_{out})+2 \delta m_n-3\delta m_{n'})\nonumber\\&\times&G^{jb}\proj{n}G^{lc}\proj{n'}G^{lc}\proj{n}G^{ia}\\
&=& i\lepspi \left(\gAoF \right)^4\frac 13 k_1^i k_2^j\left( (k_1^0+k_2^0) G^{jb} \hat G^2G^{ia} +[\delta \hat m, G^{jb}] \hat G^2 G^{ia}\right.\nonumber\\
&-&\left.G^{jb} \hat G^2[\delta \hat m,G^{ia}]+3 G^{ib}[[\delta \hat m, G^{lc}],G^{lc}]G\nonumber
^{ia}\right)\eea

{Diagrams $D_2$ UV:}

\bea
i T_{D_2}^{ba^{UV}}{\rm (no-pole)}&=&-i\lepspi \left(\gAoF \right)^4\frac 13 k_1^i k_2^j\nonumber\\
&\times&\sum_{n,n',n''}
\left( G^{jb}\proj{n}G^{lc}\proj{n''}G^{ia}\proj{n'} G^{lc}(k_1^0+2p_1^0+\delta m_n-\delta m_{n'}-2\delta m_{n''})\right.\nonumber\\
&+& \left. G^{lc}\proj{n'}G^{jb}\proj{n''}G^{lc}\proj{n}G^{ia}(k_2^0+2p_2^0+\delta m_n-\delta m_{n'}-2 \delta m_{n''})\right)\\
&=& -i\lepspi \left(\gAoF \right)^4\frac 13 k_1^i k_2^j
\left(k_1^0 G^{jb} G^{lc} G^{ia} G^{lc}+k_2^0 G^{lc} G^{jb} G^{lc} G^{ia}\right.\nonumber\\
&-& 2 G^{jb}G^{lc}G^{ia}[\delta \hat m,G^{lc}]+G^{jb}[\delta \hat m,G^{lc}]G^{ia}G^{lc}-G^{jb}G^{lc}[\delta \hat m,G^{ia}]G^{lc}\nonumber\\
&+& \left. 2[\delta \hat m,G^{lc}] G^{jb} G^{lc}G^{ia}+G^{lc}[\delta \hat m,G^{jb}]G^{lc}G^{ia}-G^{lc}G^{jb}[\delta \hat m,G^{lc}]G^{ia}\right)\nonumber
\eea

{Diagrams $D_4$ UV:}

\bea
i T_{D_4}^{ba^{UV}}&=&i\lepspi \left(\gAoF \right)^4\frac 13 k_1^i k_2^j(k_1^0+k_2^0+3p_1^0+3p_2^0-2\delta m_n-2 \delta m_{n'}-2\delta m_{n''})\nonumber\\
&\times&
G^{lc}\proj{n''}G^{jb}\proj{n'}G^{ia}\proj{n} G^{lc}\nonumber\\
&=&i\lepspi \left(\gAoF \right)^4\frac 13 k_1^i k_2^j\left( (k_1^0+k_2^0)G^{lc}G^{jb}G^{ia}G^{lc}+ 3 [\delta \hat m,G^{lc}]G^{jb} G^{ia} G^{lc}\right.\nonumber\\
&-&3G^{lc} G^{jb} G^{ia}[\delta \hat m,G^{lc}]   
 +  \left. G^{lc}[\delta \hat m,G^{jb}]G^{ia}G^{lc}-G^{lc}G^{jb}[\delta \hat m,G^{ia}]G^{lc}\right)
\eea
\end{widetext}

The added UV divergencies of the no-pole terms of the diagrams $D_1$, $D_2$ and $D_4$ give:
\begin{widetext}
\bea
\label{UVgA4}
i  T^{ba^{UV}}{\rm (no-pole)}(\gA^4)&=&i\lepspi \left(\gAoF \right)^4\frac 13 k_1^i k_2^j
\left(-(k_1^0+k_2^0)[G^{lc},G^{jb}][G^{lc},G^{ia}]\right.  \\
&+& \left. 2 \frac{C_{\rm HF}}{N_c}\left( [G^{jb},[\delta \hat m,G^{lc}]][G^{lc},G^{ia}]-[G^{lc},G^{jb}][G^{ia},[\delta \hat m,G^{lc}]]\right)\right) \nonumber
\eea
\end{widetext}
\subsubsection{\rm Projection of UV divergence onto basis of SF operators}

To the diagrams $D_1$ through $D_4$,  the crossed diagrams must be added,  which are obtained by the  prescription: $k_1\leftrightarrow -k_2$,  and $a\leftrightarrow b$. With this,  the sum of the no-pole terms proportional to $\gA^4$ gives the UV divergence in terms of bases operators in Table \ref{SF-basis}:
\begin{widetext}
\bea
i  T_{\rm no-pole}^{ba^{UV}}(\gA^4)&=&i\lepspi \left(\gAoF \right)^4\frac 16 k_1^i k_2^j\left(\frac{i}{8} (k_1^0+k_2^0)\Big( \delta^{ij}\eps^{bac}I^c+\delta^{ab} \eps^{jik} S^k \Big) \right.\nonumber\\
&+&      \frac{C_{\rm HF}}{N_c}\Big(  \delta^{ab} S^i S^j\arrowvert_{J=2}+\delta^{ij} I^a I^b\arrowvert_{I=2} -\frac 12 \eps^{jik}\eps^{bac}(N_c(N_c+2)G^{kc}-3 S^k I^c)\nonumber\\
&-&\left.   3 \;G^{jb}G^{ia}\arrowvert_{J=I=2}\Big)\vphantom{\frac1 1}\right),
\label{pUVgA4}
\eea
\end{widetext}
obtained from  Eqn.(\ref{UVgA4}) by making use of the relations:
\begin{widetext}
	\bea
	[G^{lc},G^{jb}][G^{lc},G^{ia}]\! &=&\! -\frac 1{16}\Big( \frac 43 \delta^{ij}\delta^{ab}\hat S^2+\frac i 2 \delta^{ab} \eps^{jik}S^k+\frac i2 \delta^{ij} \eps^{abc}I^c+ 2 \eps^{jik} \eps^{abc}S^k I^c \nonumber\\
	&-& \delta^{ab} S^iS^j\arrowvert_{J=2}-\delta^{ij} I^a I^b\arrowvert_{I=2}\Big)\nonumber\\
	\!\,[G^{jb},[\delta \hat m,G^{lc}]][G^{lc},G^{ia}] \!&-&\! [G^{lc},G^{jb}][G^{ia},[\delta \hat m,G^{lc}]]=\nonumber\\
	&&\!\!\!\frac{C_{\rm HF}}{N_c}\left(\frac{i}{48}(3N_c(N_c+4)+6-28\hat S^2)(\delta^{ab}\eps^{jik} S^k+\delta^{ij}\eps^{bac}I^c))\right. \nonumber\\
	\!&-&\! \frac{1}{8}\eps^{jik}\eps^{bac}(N_c(N_c+2)G^{kc}-3 S^kI^c)\nonumber\\
	&+&\frac 14(\delta^{ab} S^iS^j\arrowvert_{J=2}+\delta^{ij} I^a I^b\arrowvert_{I=2})\nonumber\\
	\!&+&\! \left.\frac i 2(\eps^{jik} S^k I^a I^b\arrowvert_{I=2}+\eps^{bac} I^c S^iS^j\arrowvert_{J=2})-3 G^{jb}G^{ia}\arrowvert_{J=I=2}\right).
	\label{UVgA4red}
	\eea
\end{widetext}
These UV divergencies give contributions that are at most $\ord{N_c^0}$, and thus consistent with the $1/N_c$ expansion of the scattering amplitudes. They are also independent of $M_\pi$. 


\subsection{Diagrams $\boldsymbol{\propto \mathring g_A^2}$}

The diagrams $D_{6,\,7,\,8}$ vanish identically. 

{Diagrams $D_5$:} 

These diagrams give  only single pole contributions,   correcting   the $\pi$-baryon coupling, and thus are finally included in the pole contributions Eqn.(\ref{NNLOpoleamplitude}).
\begin{widetext}
\bea
i T^{ba}_{D_5}&=& -i \left(\frac{\mathring g_A}{F_\pi^2} \right)^2\frac 23 k_1^i k_2^j\; \Delta(M_\pi)\sum_{n}\; G^{jb} \proj{n} G^{ia} \frac{1}{p_1^0+k_1^0-\delta m_n}
\eea
where:
\beq
\Delta(M_\pi)=\int \widetilde{d^d k}\frac{i}{k^2-M_\pi^2}=-\frac{M_pi^2}{(4\pi)^2}\Big(\lambda_\eps-\log \frac{M_\pi^2}{\mu^2}\Big)
\eeq

{Diagrams $D_9$:}

\bea
i T^{ba}_{D_9}&=&  \left(\frac{\mathring g_A}{F_\pi^2} \right)^2 \frac 16 \sum_{n}\; G^{kc}\proj{n}G^{kd}(\delta^{ac}\delta^{bd}+\delta^{ad}\delta^{bc}-2\delta^{ab}\delta^{cd})\nonumber\\
&\times& ({\cal{I}}(\delta m_n-p_1^0,M_\pi)+{\cal{I}}(\delta m_n-p_2^0,M_\pi))
\eea

{Diagram $D_{10}$:} 

this diagram  is more compactly expressed in terms of the integrals defined earlier in Eqns.(\ref{Jintegrals1},\ref{Jintegrals2}), and reads:

\bea
i T^{ba}_{D_{10}}&=& -i \left(\frac{\mathring g_A}{F_\pi^2} \right)^2\frac 13 \sum_{n}G^{ld} \proj{n} G^{jc} \frac{1}{(4\pi)^{d/2}}\left(\frac 12 \delta^{jl}(\delta^{ac}\delta^{bd}+\delta^{ad}\delta^{bc}-2\delta^{ab}\delta^{cd})\right.\nonumber\\
&\times&\Gamma(1-\frac d2)(J(0,1-\frac d 2,C_0^+,1,\lambda_0^+)+J(0,1-\frac d 2,C_0^-,1,\lambda_0^-))\nonumber\\
&+& 6\int_0^1 d\alpha\left(((t-M_\pi^2)\delta^{ab}\delta^{cd}+\frac 12(2M_\pi^2-t)(\delta^{ac}\delta^{bd}+\delta^{ad}\delta^{bc} ))\right.\nonumber\\
&\times& (\Gamma(3-\frac d2)J(0,3-\frac d 2,C_0^1,1,\lambda_0^1)-\frac 12\delta^{jl}\Gamma(2-\frac d2)J(0,2-\frac d 2,C_0^1,1,\lambda_0^1))\nonumber\\
&+&(\delta^{ac}\delta^{bd}-\delta^{ad}\delta^{bc})\left(\frac 12(\alpha q^l(k_1+k_2)^j+(\alpha-1)q^j(k_1+k_2)^l )\Gamma(2-\frac d2)J(0,2-\frac d 2,C_0^1,1,\lambda_0^1)\right.\nonumber\\
&+& \alpha(1-\alpha) q^j q^l (k_1^0+k_2^0)\Gamma(3-\frac d2)J(1,3-\frac d 2,C_0^1,1,\lambda_0^1)\nonumber\\
&-& \left.\left.\left.\frac 12 \delta^{jl} (k_1^0+k_2^0)\Gamma(2-\frac d2)J(1,2-\frac d 2,C_0^1,1,\lambda_0^1) \right) \right) \right)
\eea
where:
\bea 
d&=&4-2\eps,\quad q= k_1-k_2,\quad t= q^2\nonumber\\
 C_0^\pm&=&M_\pi^2-\lambda_0^\pm,\quad
\lambda_0^\pm= \frac 12(p_1^0+p_2^0\pm q^0)-\delta m_n\nonumber\\
C_0^1&=&M_\pi^2+\alpha(\alpha-1)q^2-\lambda_0^1,\quad
\lambda_0^1=\frac 12(p_1^0+p_2^0)+(\frac 12 -\alpha) q^0-\delta m_n.
\eea

{Diagram $D_{11}$:}

\bea
i T^{ba}_{D_{11}}&=& i \left(\frac{\mathring g_A}{F_\pi^2} \right)^2 \sum_{n,n'}G^{kc}\proj{n'}(-\frac 12 (k_1^0+k_2^0)\eps^{bac} I^c +2i c_1 N_c \frac{M_\pi^2}{\Lambda}\delta^{ab})\proj{n}G^{kd} \nonumber\\
&\times& \frac{1}{p_2^0-p_1^0-\delta m_{n'}+\delta m_n} ({\cal{I}}(\delta m_{n'}-p_2^0,M_\pi)-{\cal{I}}(\delta m_{n}-p_1^0,M_\pi))
\eea 

{Diagram $D_{12}$:}

\bea
i T^{ba}_{D_{12}}&=& \frac 12 \frac{1}{F_\pi^2} \Big\{\frac12 (k_1^0+k_2^0)\eps^{bac} I^c +2i c_1 N_c \frac{M_\pi^2}{\Lambda}\delta^{ab} ,\; \delta \hat Z\Big\}
\eea
\end{widetext}
where $\delta Z$ is evaluated on-shell for the in and out baryons, which in this case have the same spin
.
\subsubsection{\rm UV divergencies of the no-pole terms}
The UV divergent pieces of no-pole terms of the diagrams proportional to $  \mathring g_A^2$, evaluated for the external baryons on mass shell are as follows:.

\begin{widetext}
{Diagram $D_9$ UV:}

\bea
iT_{D_{9}}^{ba^{UV}}&=&-i\frac{\lambda_\eps}{(4\pi)^2}\left(\frac{ \gA }{  
	F_\pi^2}\right)^2 \sum_{n}G^{ie} \proj{n}  G^{ic}\,(\delta^{ae} \delta^{bc}+\delta^{ac} \delta^{be}-2
\delta^{ab} \delta^{ce}) \\
&\times&\frac{1}{16}\left(3 M_\pi^2 ({(p_1^0-\delta m_n)+(p_2^0-\delta m_n)})-2 \left({(p_1^0-\delta m_n)}^3+{(p_2^0-\delta m_n)}^3\right)\right).\nonumber 
\eea
This diagram projects onto $J=0, \;I=0,2$, implying  that $p_1^0=p_2^0=\delta m_{in}$, and  yielding:

\bea
iT_{D_{9}}^{ba^{UV}}(J=I=0)&=&i\frac{\lambda_\eps}{(4\pi)^2}\left(\frac{ \gA }{  
F_\pi^2}\right)^2 \delta^{ab} \frac 29 \frac{  C_{\rm HF}}{N_c}  \left(M_\pi^2 \Big(5 \hat S^2-\frac{3}{4} N_c
(N_c+4)\Big)\right.\nonumber\\
&+& \left. \left(\frac{C_{\rm HF}}{N_c}\right)^2
\Big(\frac{2}{3} (5 N_c (N_c+4)-24) \hat S^2+2 N_c (N_c+4)-\frac{56
\hat S^4}{3}\Big)\right) \\
iT_{D_{9}}^{ba^{UV}}(J=0,I=2)&=&i\frac{\lambda_\eps}{(4\pi)^2}\left(\frac{ \gA }{  
F_\pi^2}\right)^{\!2}\!\! I^aI^b\arrowvert_{I=2}\; \frac 13 \frac{C_{\rm HF}}{N_c}  \left(M_\pi^2+\left(\frac{C_{\rm HF}}{N_c}\right)^2
\Big(\frac{2}{3} N_c (N_c+4)-\frac{16
\hat S^2}{3}\Big) \right). \nonumber
\eea

{Diagram $D_{10}$ UV:} 

upon projecting the diagram onto t-channel $(J,I)$, the corresponding UV divergent amplitudes are the following:
\bea
iT_{D_{10}}^{ba^{UV}}(J=I=0)&=	 & \frac{i}{54 } \frac{\lambda_\eps}{(4\pi)^2}\left(\frac{\gA}{F_\pi^2}\right)^2  \sum_{n}G^{ic}\proj{n}G^{ic } \delta^{ab}
(p_1^0+p_2^0-2 \delta m_n)\nonumber\\
&\times& \left(6 (k_1^0-k_2^0)^2-21 M_\pi^2+2 \left((p_1^0+p_2^0-2 \delta m_n)^2+9 t\right)\right)  \nonumber\\
&=&i \frac{\lambda_\eps}{(4\pi)^2}\left(\frac{\gA}{F_\pi^2}\right)^2  \frac 19 \delta^{ab}  \frac{  C_{\rm HF}}{N_c}   \left(-\frac{3}{8} N_c
(N_c+4) \left(5 M_\pi^2-12 k_1\cdot k_2\right)\right.
\nonumber\\
&+&\left(\frac{C_{\rm HF}}{N_c}\right)^2
\Big(-\frac{4}{3} (5 N_c (N_c+4)-24) \hat S^2-4 N_c
(N_c+4)+\frac{112 }{3}\hat S^4\Big)\nonumber\\
&+&\left.\hat S^2 \Big(\frac{25
}{2}M_\pi^2-30 k_1\cdot k_2\Big)
\right) 
\eea
\bea  
iT_{D_{10}}^{ba^{UV}}(J=0,\, I=1)&=	 & -\frac{i}{18 } \frac{\lambda_\eps}{(4\pi)^2}\left(\frac{\gA}{F_\pi^2}\right)^2  \sum_{n} G^{ie}\proj{n}G^{ic }  \frac 12(\delta^{ae}\delta^{bc}-\delta^{ac}\delta^{be})\nonumber\\
&\times& \left((k_1^0+k_2^0) \left(3 \left((k_1^0-k_2^0)^2-6 M_\pi^2+3 (p_1^0+p_2^0-2 \delta m_n)^2+t\right)-2  (\vec k1-\vec k2 )^2\right)\right.\nonumber\\
&+&\left. 2 (k_1^0-k_2^0) (\vec k_1^2-\vec k_2^2) \right)
\nonumber\\
&=& \frac{\lambda_\eps }{(4 \pi)^2 } \left(\frac{  \gA}{F_\pi^2}\right)^2 \frac 14 \eps^{bac}I^c    (k_1^0+k_2^0) 
\bigg(\frac{1}{6}
\left(3 k_1\cdot k_2+(\vec k_1-\vec k_2)^2+6 M_\pi^2\right) \nonumber\\
&+& \left(\frac{C_{\rm HF}}{N_c}\right)^2 (3 N_c (N_c+4)-20 \hat S^2)\bigg) 
\eea
\bea  
iT_{D_{10}}^{ba^{UV}}(J=0,\, I=2)&=	 & -\frac{i}{18 } \frac{\lambda_\eps}{(4\pi)^2}\left(\frac{\gA}{F_\pi^2}\right)^2  \sum_{n} G^{ie}\proj{n}G^{ic }  \frac 12 (\delta^{ae} \delta^{bc}+\delta^{be} \delta^{ac}-\frac 23 \delta^{ab} \delta^{ce})\nonumber\\
&\times&(p_1^0+p_2^0-2 \delta m_n) \left(3 (k_1^0-k_2^0)^2-24 M_\pi^2+(p_1^0+p_2^0-2 \delta m_n)^2+9
t\right)    \nonumber\\
&=&i \frac{\lambda_\eps}{(4\pi)^2}\left(\frac{\gA}{F_\pi^2}\right)^2    I^aI^b\arrowvert_{I=2} \;\frac 19\frac{  C_{\rm HF}}{N_c}    \Big(-3 \left(3 k_1\cdot k_2+M_\pi^2\right)\phantom{\frac 1 1} \nonumber\\
&+& \left(\frac{C_{\rm HF}}{N_c}\right)^2 (16
\hat S^2-2 N_c (N_c+4))\Big) 
\eea
\bea
iT_{D_{10}}^{ba^{UV}}(J=I=1)&=	 &-i \frac{\lambda_\eps}{(4\pi)^2}\left(\frac{\gA}{F_\pi^2}\right)^2   \sum_{n} G^{le}\proj{n}G^{jc }  \frac 12(\delta^{ae}\delta^{bc}-\delta^{ac}\delta^{be})\nonumber\\
&\times& (p_1^0+p_2^0-2 \delta m_n)  
((\vec k_1^j-\vec k_2^j ) (\vec k_1^l+\vec k_2^l)) \arrowvert_{J=1} \nonumber\\
&=& i\frac{\lambda_\eps }{(4 \pi)^2 } \left(\frac{  \gA}{F_\pi^2}\right)^2 \eps^{bac} \eps^{jil}\frac{  C_{\rm HF}}{N_c}  \frac 12 k_1^i k_2^j\; (G^{lc}
(N_c+2)-3 I^c S^l) 
\eea

{Diagram $D_{11}$ UV:}

\bea
iT_{D_{11}}^{ba^{UV}}&=	 &-\frac{\lambda_\eps}{(4\pi)^2}\left(\frac{\gA  }{    F_\pi^2  }\right)^2 \frac 16   \sum_{n}\left(3 M_\pi^2-2 \left((p_1^0-\delta m_n) (p_2^0-\delta m_n)+(p_1^0-\delta m_n)^2+(p_2^0-\delta m_n)^2\right)\right) \nonumber\\
&\times&\left( \eps^{abc}  (k_1^0+k_2^0) 
G^{ie}\proj{n}I^cG^{ie}-4 i  \delta^{ab}c_1N_c \frac{M_\pi^2 }{\Lambda}
G^{ie}\proj{n}G^{ie}\right)
\eea
\bea
iT_{D_{11}}^{ba^{UV}}(J=I=0)&=	 &i\frac{\lambda_\eps }{(4 \pi)^2 } \left(\frac{  \gA}{F_\pi^2}\right)^2 \delta^{ab} \text{c1} N_c  \frac{M_\pi^2}{\Lambda}  
\bigg(M_\pi^2 \Big(\frac{3}{8} N_c
(N_c+4)-\hat S^2\Big) \nonumber\\
&+& \left(\frac{C_{\rm HF}}{N_c}\right)^2 \left(-2 (N_c-2) (N_c+6) \hat S^2-3 N_c
(N_c+4)+8 \hat S^4\right) \bigg)  \nonumber\\
iT_{D_{11}}^{ba^{UV}}(J=0,\,I=1)&=  	 &\frac{\lambda_\eps }{(4 \pi)^2 } \left(\frac{  \gA}{F_\pi^2}\right)^2 \frac 14  \eps^{bac}  I^c   (k_1^0+k_2^0) \left(M_\pi^2 \Big(\frac{3}{8} (N_c
(N_c+4)-4)-\hat S^2\Big)\right.\nonumber\\
&+&\left.\left(\frac{C_{\rm HF}}{N_c}\right)^2
\left(-2 N_c(N_c+4) (\hat S^2+3)+4 \hat S^2 (2
\hat S^2+11)\right) \right)
\eea

{Diagram $D_{12}$ UV:}

\bea
iT_{D_{12}}^{ba^{UV}}&=	 & \frac{\lambda_\eps}{ (4  \pi)^2}   \left(\frac{\gA}{F_\pi^2}\right)^2 \frac 12   \sum_{n}G^{id}\proj{n} G^{id}(M_\pi^2 - (p_1^0-\delta m_n)^2 - 
(p_2^0-\delta m_n)^2) \nonumber\\
&\times& (-4 i \delta^{ab}c_1 N_c \frac{M_\pi^2}{\Lambda}  + \eps^{abc}
I^c (k_1^0 + k_2^0)   )
\eea

\bea
iT_{D_{12}}^{ba^{UV}}(J=I=0)&=	 &i\frac{\lambda_\eps }{(4 \pi)^2 } \left(\frac{  \gA}{F_\pi^2}\right)^2 \delta^{ab} \text{c1} N_c  \frac{M_\pi^2}{\Lambda}  
\bigg(M_\pi^2 \Big(\hat S^2-\frac{3}{8}
N_c (N_c+4)\Big) \nonumber\\
&+& \left(\frac{C_{\rm HF}}{N_c}\right)^2 \left(2 (N_c-2) (N_c+6) \hat S^2+3 N_c
(N_c+4)-8 \hat S^4\right) \bigg)\nonumber\\
iT_{D_{12}}^{ba^{UV}}(J=0,\,I=1)&=	 & \frac{\lambda_\eps }{(4 \pi)^2 } \left(\frac{  \gA}{F_\pi^2}\right)^2 \frac 14 \eps^{bac}I^c  (k_1^0+k_2^0) \left( M_\pi^2 \Big(\hat S^2-\frac{3}{8} N_c
(N_c+4)\Big)\right. \nonumber\\&+&\left.\left(  \frac{C_{\rm HF}}{N_c}\right)^2
\left(2 (N_c-2) (N_c+6) \hat S^2+3 N_c (N_c+4)-8
\hat S^4\right)\right)
\label{gA2UVJI}
\eea
Adding up  the diagrams, the $\gA^2$ UV divergencies yield:
\bea
iT^{ba^{UV}}(\gA^2)&=	 &i\frac{\lambda_\eps }{(4 \pi)^2 } \left(\frac{  \gA}{F_\pi^2}\right)^2 \frac 14 \left(2\frac{  C_{\rm HF}}{N_c} \eps^{bac} \eps^{jil}k_1^i k_2^j
(G^{lc} (N_c+2)-3 I^c S^l)\right.\nonumber\\
&+&\left.\frac{i}{2} \eps^{bac} I^c (k_1^0+k_2^0)
\Big(-k_1\cdot k_2-\frac{1}{3}
(\vec k_1-\vec k_2)^2+M_\pi^2\Big)\right.
\nonumber\\
&+&\left.\frac{C_{\rm HF}}{N_c}
\delta^{ab} \Big(k_1\cdot k_2 \Big(2 N_c (N_c+4)-\frac{40
\hat S^2}{3}\Big)+M_\pi^2 \Big(\hat S^2-\frac{3}{2} N_c
(N_c+4)\Big)\Big)\right.\nonumber\\
&-&\left.4\frac{  C_{\rm HF}}{N_c} I^aI^b\arrowvert_{I=2}\;
k_1\cdot k_2\right).
\label{UVgA2}
\eea
\end{widetext}
In the $J=0$ terms $k_1^0=k_2^0$, which in the term $(J=0,I=1)$ gives the simplification: $(-k_1\cdot k_2-\frac{1}{3}
(\vec k_1-\vec k_2)^2+M_\pi^2)\to\frac 16 t$.

\subsection{Diagrams $\boldsymbol{\propto \mathring g_A^0}$}
In these diagrams the baryon flowing through it has a fixed spin, thus in the CM frame $k_1^0=k_2^0$, and each diagram satisfies large $N_c$ consistency. The results are as follows:
 
\begin{widetext}
{Diagram $D_{13}$:}
\bea
i T^{ba}_{D_{13}}&=&-\frac{1}{(4\pi)^2}\frac 5{F_\pi^4} M_\pi^2\Big(\lambda_\eps+1-\log\frac{M_\pi^2}{\mu^2}\Big)\left(\frac{1}{24}\eps^{bac}I^c (k_1^0+k_2^0)+\frac i 3 \delta^{ab}c_1    N_c \frac{M_\pi^2}{\Lambda}\right)
\eea

{Diagram $D_{14}$:}
 
\bea
	i T^{ba}_{D_{14}}&=& \frac{1}{F_\pi^4} \frac{1}{(4\pi)^2}\bigg(i c_1 N_c \frac{M_\pi^2}{\Lambda}     \delta^{a b}   \Big(\Big(\frac{7}{3} M_\pi^2 -2 t\Big)\Big(\lambda_ \eps-\log \frac{M_\pi^2}{\mu^2}\Big)+\frac{10 M_\pi^2}{3}-4 t  \nonumber\\
	&-&2   \sqrt{1-\frac{4 M_\pi^2}{t}}\left(M_\pi^2-2 t\right) \arctan  \frac{\sqrt{t}}{\sqrt{-4 M_ \pi^2+t}} \Big)\nonumber\\ 
	&+&2 \eps^{bac} I^c(k_1^0+k_2^0)  \left(\Big(M_\pi^2-\frac{t}{6}\Big)\Big(\lambda_\eps-\log  \frac{M_\pi^2}{\mu^2} \Big)+ \frac{7}{3} M_\pi^2- 4 t  \right.\nonumber  \\ &+&  \frac{1}{3}\left(-4 M_\pi^2+t\right)^{3 / 2} \arctan  \frac{\sqrt{t}}{\sqrt{-4 M_\pi^2+t}}   \Big)\bigg)  
	\eea

{Diagram $D_{15}$:}
\bea
i T^{ba}_{D_{15}}&=&-\frac{1}{F_\pi^4}\left( \frac i 4\eps^{bac}I^c \left(\frac 12 \Gamma(1-\frac d2)J(0,1-\frac d2,C_0,1,\lambda_0)\right.\right.\nonumber\\
&-&  \Gamma(2-\frac d2)(k_1^0k_2^0 J(0,2-\frac d2,C_0,1,\lambda_0)+(k_1^0+k_2^0)J(1,2-\frac d2,C_0,1,\lambda_0) \nonumber\\
&+&\left.J(2,2-\frac d2,C_0,1,\lambda_0)+ 4c_1 N_c \frac{M_\pi^2}{\Lambda}((k_1^0+k_2^0) J(0,2-\frac d2,C_0,1,\lambda_0)\right.\nonumber\\
&+&\left.2J(1,2-\frac d2,C_0,1,\lambda_0)))\right)\nonumber\\
&-&\left. 4 \delta^{ab}c_1^2 N_c^2 \frac{M_\pi^4}{\Lambda^2} \Gamma(2-\frac d2) J(0,2-\frac d2,C_0,1,\lambda_0)\right),
\eea
where $C_0=M_\pi^2-\lambda_0^2$, and $\lambda_0=k_1^0=k_2^0$. Adding the   crossed diagram and, in order to simplify the result, using explicitly that $k_1^0=k_2^0$ in the CM frame,  yields:
\bea
i T^{ba}_{D_{15}+{\rm crossed}}&=&\frac{1}{(4\pi)^2}\frac{1}{F_\pi^4}
\left( -i\delta^{ab}8\pi\sqrt{M_\pi^2-k_1^{0^2}}
\Big(  \frac 13 k_1^0\hat S^2+2 c_1^2 N_c^2 \frac{M_\pi^4}{\Lambda^2}  \Big)\right.\nonumber\\
&+& \eps^{bac}I^ck_1^0\Big( \big(\frac 34 M_\pi^2-2 k_1^{0^2}\big)\Big(\lambda_\eps+1-\log\frac{M_\pi^2}{\mu^2}\Big)-2 k_1^{0^2} 
\nonumber\\
&+& 4\sqrt{M_\pi^2-k_1^{0^2}} \Big( -4\pi  c_1 N_c \frac{M_\pi^2}{\Lambda} +k_1^0\arctan\frac{k_1^0}{\sqrt{M_\pi^2-k_1^{0^2}} } \Big)\Big) \nonumber\\
&+& \left. i 4\pi k_1^{0^2} \sqrt{M_\pi^2-k_1^{0^2}} I^aI^b\arrowvert_{I=2} \right) 
\eea

\subsubsection{\rm UV divergencies}
{Diagram $D_{13}$ UV:}
\bea
i T^{ba^{UV}}_{D_{13}}&=&-\lepspi\frac 5{F_\pi^4} M_\pi^2(\frac{1}{32}\eps^{bac}I^c (k_1^0+k_2^0)+\frac i 3 \delta^{ab}c_1 N_c \frac{M_\pi^2}{\Lambda})
\eea

{Diagram $D_{14}$ UV:}
\bea
i T^{ba^{UV}}_{D_{14}}&=&\lepspi\frac 1{F_\pi^4}\frac {1}{12}\left( 4i   c_1 N_c \frac{M_\pi^2}{\Lambda} \delta^{ab}  (7M_\pi^2-6t) +\eps^{bac}I^c (k_1^0+k_2^0)  (6M_\pi^2-t) \right) 
\eea

{Diagram $D_{15}$ UV:} 

Adding the crossed diagram to $D_{15}$ yields:
\bea
i T^{ba^{UV}}_{D_{15}+{\rm crossed}}&=&-\lepspi\frac 1{F_\pi^4} \eps^{bac} I^c \frac 14 (k_1^0+k_2^0)\Big((k_1^0+k_2^0)^2-\frac 32 M_\pi^2\Big)
\eea

Adding up the diagrams, the   UV divergences  proportional to $\gA^0$ become:
\bea
i  T^{ba^{UV}}(\gA^0)&=&\lepspi \frac{1}{12 F_\pi^4}\left(i \delta^{ab} 8 c_1 N_c \frac{M_\pi^2}{\Lambda}(M_\pi^2-3t)\right.\nonumber\\
&+&\left. \phantom{\frac{1}{1}}\eps^{bac}I^c (k_1^0+k_2^0)(3(k_1^0+k_2^0)^2-8 M_\pi^2+t) \right)
\label{UVgA0}
\eea


\subsection{${\mathbf{\beta}}$-functions}

\begin{center}
\begin{table}
\begin{tabular}{|c|c|c|c|c|c|}
	\hline\hline
\;\;	LEC\;\; & $\beta_{f_{JI}}/\left(\frac{\gA^4}{F_\pi^2}\right)$ &\;\; LEC\;\; &$\beta_{g_{JI}}/\left(\Lambda\frac{\gA^2}{F_\pi^2}\right)$  &LEC &$\beta_{h_{JI}}/\left(\frac{\Lambda^2}{F_\pi^2}\right)$  \\
	\hline
$f^{(1)}_{10}$	 & $\frac{\Lambda^2}{96} $&$g^{(1)}_{00}$ & $-\frac{3}{32}(N_c+4)C_{\rm HF}$ & $h_{00}^{(1)}+h_{00}^{(2)}$& $-\frac{5}{24}c_1$\\

$f^{(1)}_{01}$ &  $\frac{\Lambda^2}{96} $&$g^{(2)}_{00}$ &$\frac{1}{16}   C_{\rm HF}$ & $h_{00}^{(3)}$& $-\frac{1}{2}c_1$\\
	 
$f^{(1)}_{11}$ & $ \frac{N_c+2}{12 N_c}\Lambda C_{\rm HF}$ &$g^{(3)}_{00}$ &$-\frac{1}{4}(N_c+4)C_{\rm HF}$ &$h_{01}^{(1)}$ & $\frac 1 2$\\
	 
$f^{(2)}_{11}$ & $-\frac{1}{4}\Lambda C_{\rm HF}$ &$g^{(4)}_{00}$ & $\frac{5}{3} C_{\rm HF}$&$h_{01}^{(2)}$ & $\frac 1{12}$\\
	 
$f_{02}$ &  $-\frac{1}{6}\Lambda C_{\rm HF}$  & $g^{(1)}_{01}$& $-\frac{1}{48}\Lambda$ &$h_{01}^{(3)}$ & $\frac 1 4$ \\
	 
$f_{20}$ &  $-\frac{1}{6}\Lambda C_{\rm HF}$& $g^{(2)}_{01}$& $-\frac{1}{48}\Lambda$ & & \\
	 
 $f_{22}$ &  $ \Lambda C_{\rm HF}$& $g^{(1)}_{02}$&$\frac{1}{2N_c}C_{\rm HF}$  & & \\
	 
& &$g^{(1)}_{11}$ & $\frac {N_c+2}{4N_c} C_{\rm HF}$& & \\
&&$g^{(2)}_{11}$&$-\frac 34 C_{\rm HF}$&&\\
	\hline\hline
\end{tabular}
\caption{$\beta$ functions of the NNLO Lagrangians Eqn.(\ref{UV-CTs4},\ref{UV-CTs2},\ref{UV-CTs0}).}
\label{betafunctionstable}
\end{table}
\end{center}
\end{widetext}
The $\beta$ functions corresponding to the LECs of the CT Lagrangians Eqn.(\ref{UV-CTs4},\ref{UV-CTs2},\ref{UV-CTs0})  are given in  Table \ref{betafunctionstable}. The definition of the $\beta$-function for a given LEC $X$ is the following:
\beq
X\equiv X(\mu)+\beta_X \frac{\lambda_\eps}{(4\pi)^2},
\eeq
where $X(\mu)$ is the renormalized LEC at the renormalization scale $\mu$. For the Lagrangian Eqn.(\ref{Finite-CTs}) the $\beta$-functions obviously vanish.

\subsection{Counterterm contributions to  the  amplitudes.}

The   contributions from the   Lagrangians Eqns.(\ref{UV-CTs4},\ref{UV-CTs2},\ref{UV-CTs0},\ref{Finite-CTs}) to the scattering amplitudes for definite t-channel $(J,I)$ are the following:
\begin{widetext}
\bea
i T_{CT}^{ba}(J=I=0)&= &  i2 \frac{\delta^{a b}}{F_\pi^2} \left(-2\frac{g_{00}^{(1)}}{\Lambda}   M_\pi^2-2\frac{g_{00}^{(2)}}{\Lambda}   M_\pi^2 \frac{ \hat S^2}{N_c}\right. \nonumber\\  & -&8\frac{\left(h_{00}^{(1)}+h_{00}^{(2)}\right)}{\Lambda^3} N_c  M_\pi^4+\frac{g_{00}^{(3)}}{\Lambda} k_1 \cdot k_2 \nonumber\\  & +&\frac{g_{00}^{(4)}}{\Lambda} k_1 \cdot k_2 \frac{ \hat S^2}{N_c}+2\frac{h_{00}^{(3)}}{\Lambda^3} N_c   M_\pi^2 k_1 \cdot k_2 \nonumber\\  & +&\left. \frac{1}{2} \frac{l_{00}^{(1)}}{\Lambda^2}(k_1^i k_2^0+k_1^0 k_2^i)(p_1^i+p_2^i)   -2\frac{l_{00}^{(2)}}{\Lambda^2}   M_\pi^2  (p_1^0+p_2^0)\right)
\label{piNCTs00}
\eea
\bea
i T_{CT}^{ba}(J=0,I=1)& = &\frac{1}{F_\pi^2} \eps^{a b c} I^c\left(2\frac{\tilde f_{01}^{(1)}}{\Lambda^2}    k_1 \cdot k_2(k_1^0+k_2^0)    +2 \frac{\tilde  g_{01}^{(1)}}{\Lambda^2} M_\pi^2(k_1^0+k_2^0)\right.\nonumber\\
& -&\frac{l_{01}^{(1)}}{2 m_0}(\vec k_1+\vec k_2 ) \cdot(\vec p_1+\vec p_2)   -\frac{l_{01}^{(2)}}{\Lambda^2}(k_1^i+k_2^i)(k_1^i k_2^0+k_1^0 k_2^i) \nonumber\\
& +&\left.\frac{l_{01}^{(3)}}{\Lambda^2}(k_1^0 \vec k_2^2+k_2^0\vec k_1^2-\vec k_1 \cdot \vec k_2 (k_1^0 +k_2^0 ))\right)
\label{piNCTs01}
\eea
\bea
i T_{CT}^{ba}(J=1,I=0)& = & \frac{\eps^{i j k} S^k}{F_\pi^2}\left(-2\frac{f_{10}^{(1)}}{\Lambda^2}(k_1^0+k_2^0)  k_1^i k_2^j\right. \nonumber\\
& -&\frac{l_{10}^{(1)}}{\Lambda^2}(k_1^j+k_2^j)(k_1^i k_2^0-k_2^i k_1^0) \nonumber\\
& +&\left. \frac{l_{10}^{(2)}}{\Lambda^2}(k_1^j-k_2^j)(k_1^i k_2^0+k_2^i k_1^0)\right)
\label{piNCTs10}
\eea
\bea
i T_{CT}^{ba}(J=I=1)& = & i\frac{ \eps^{i j k} \eps^{a b c}}{F_\pi^2}\left(2\frac{\tilde f_{11}^{(1)}}{\Lambda}   k_1^i k_2^j G^{k c}\right.  +2\frac{\tilde f_{11}^{(2)}}{\Lambda} \frac{1}{N_c}   k_1^i k_2^j S^k I^c \nonumber\\
& +&\left.\frac{l_{11}^{(1)}}{\Lambda^2}(k_1^i k_2^0-k_2^i k_1^0)(p_1^j+p_2^j) G^{kc}+\frac{l_{11}^{(2)}}{\Lambda^2} 2 k_1^i k_2^j(p_1^0+p_2^0)G^{kc}\right) \nonumber
\label{piNCTs11}.
\eea
\end{widetext}
These contributions for   the  $\pi N\to \pi N$ amplitudes are significantly simplified. For matrix elements between nucleon states, the generator $G^{ia}$ is replaced by: $G^{ia}\to \frac{N_c+2}{3}  S^i I^a$. In the CM frame, the CT amplitudes then become:
\begin{widetext}
\bea
i T_{CT}^{ba}(J=I=0)&=& i\frac{1}{F_\pi^2}\delta^{ab}\left(\alpha_{00}^{(1)} \frac{M_\pi^2}{\Lambda}+\alpha_{00}^{(2)} N_c \frac{M_\pi^4}{\Lambda^3}+\alpha_{00}^{(3)} \frac{k_1\cdot k_2}{\Lambda}\right.\nonumber\\
&+&\left. \alpha_{00}^{(4)} N_c\frac{M_\pi^2 k_1\cdot k_2}{\Lambda^3}+\alpha_{00}^{(5)}\frac{k^0}{\Lambda} (\vec k_1+\vec k_2)^2\right)\nonumber\\
i T_{CT}^{ba}(J=0,I=1)&=&\frac{1}{F_\pi^2} \eps^{bac} I^c\left(\alpha_{01}^{(1)}M_\pi^2 \frac{k^0}{\Lambda^2}+\alpha_{01}^{(2)}\frac{k^0}{\Lambda^2} k_1\cdot k_2+\alpha_{01}^{(3)}\frac{1}{m_0} (\vec k_1+\vec k_2)^2\right.\nonumber\\
&+&\left. \alpha_{01}^{(4)}\frac{k^0}{\Lambda^2} (\vec k_1+\vec k_2)^2+\alpha_{01}^{(5)}\frac{k^0}{\Lambda^2} (\vec k_1-\vec k_2)^2\right)\nonumber\\
i T_{CT}^{ba}(J=1,I=0)&=&\frac{1}{F_\pi^2}   \eps^{jik} S^k \alpha_{10}^{(1)} \frac{k^0}{\Lambda^2} k_1^i k_2^j \nonumber\\
i T_{CT}^{ba}(J=I=1)&=& i\frac{1}{F_\pi^2} \eps^{bac}  \eps^{jik} S^k I^c   k_1^i k_2^j \left(\alpha_{11}^{(1)} \frac{N_c}{\Lambda }  +\alpha_{11}^{(2)} \frac{k^0}{\Lambda^2 }\right),
\label{piNCTR}
\eea  
\end{widetext}
 where the LECs $\alpha_{JI}^{(n)}$ are combinations of those in Eqn.(\ref{piNCTs00},\ref{piNCTs01},\ref{piNCTs10},\ref{piNCTs11}).
 

\subsection{Reductions of spin-flavor operators in the $\pi N\to\pi N$ amplitudes}\label{SFreductionspiN}

This Appendix provides the reductions of 2-, 3-, and 4-body spin-flavor tensor operators for matrix elements between nucleon states, as needed in the calculation  of the $\pi N\to \pi N$ amplitudes. The intermediate state $\Delta_5$ of spin 5/2 is necessary for the correct general $N_c$ results, and   is obviously  absent at $N_c=3$ as shown in the corresponding entries.
\begin{widetext}
\subsubsection{\rm 2-body operators}

\begin{center}
\begin{table}[h!]
	\begin{tabular}{l|l|l|l}
		\hline\hline  $n $& \;\;\;\;$ G^{i a} \mathcal{P}_n G^{i a} $& $ \;\;\;\;\eps^{a b c} G^{i b} \mathcal{P}_n G^{i a} $&\;\;\;\; $ \eps^{i j k} \eps^{a b c} G^{j b} \mathcal{P}_n G^{i a} $\\
		\hline $N $& $ \frac{1}{16}\left(N_c+2\right)^2 $& $ -\frac{i}{12} I^c\left(N_c+2\right)^2 $& $ -\frac{1}{9} I^c S^k\left(N_c+2\right)^2 $\\
		$\Delta $& $ \frac{1}{8}\left(N_c-1\right)\left(N_c+5\right) $& $ \frac{i}{12} I^c\left(N_c-1\right)\left(N_c+5\right) $& $ -\frac{1}{18} I^c S^k\left(N_c-1\right)\left(N_c+5\right) $\\
		\hline\hline
		\end{tabular}
		\end{table}
\end{center}

\subsubsection{\rm 3-body operators}

\begin{center}
\(\begin{array}{l|l}\hline\hline
	{n} \;\;\; & \;\;\; G^{ia} \proj{n} I^c G^{ia} \\ \hline
	N \;\;\; & \;\;\; -\frac{1}{48} I^c  (N_c+2 )^2 \\
	\Delta  \;\;\; & \;\;\; \frac{5}{24} I^c  (N_c-1 )  (N_c+5 ) \;\;\;\\ \hline\hline
\end{array}\)
\end{center}

\subsubsection{\rm 4-body operators}

	\begin{center}
	\begin{table}[h!]
		\begin{tabular}{c|l|l}
			\hline\hline $n_1$   $n_2$   $n_3$ &\;\;\;\; $G^{i a} \mathcal{P}_{n_3} G^{k c} \mathcal{P}_{n_2} G^{k c} \mathcal{P}_{n_1} G^{i a}$ &\;\;\;\; $\eps^{a b c} G^{i b} \mathcal{P}_{n_3} G^{k d} \mathcal{P}_{n_2} G^{k d} \mathcal{P}_{n_1} G^{i a}$ \\
			\hline $N$   $N$  $N$ & $\frac{1}{256}\left(N_c+2\right)^4$ & $-\frac{i}{192} I^c\left(N_c+2\right)^4$ \\
			$N$   $\Delta$   $N$ & $\frac{1}{128}\left(N_c-1\right)\left(N_c+2\right)^2\left(N_c+5\right)$ & $-\frac{i}{96} I^c\left(N_c-1\right)\left(N_c+2\right)^2\left(N_c+5\right)$ \\
			$\Delta$   $N$  $\Delta$ & $\frac{1}{256}\left(N_c-1\right)^2\left(N_c+5\right)^2$ & $\frac{i}{384} I^c\left(N_c-1\right)^2\left(N_c+5\right)^2$ \\
			$\Delta$   $\Delta$  $\Delta$ & $\frac{1}{128}\left(N_c-1\right)\left(N_c+2\right)^2\left(N_c+5\right)$ & $\frac{i}{192} I^c\left(N_c-1\right)\left(N_c+2\right)^2\left(N_c+5\right)$ \\
			$\Delta$   $\Delta_5$   $\Delta$ & $\frac{3}{256}\left(N_c-3\right)\left(N_c-1\right)\left(N_c+5\right)\left(N_c+7\right)$ & $\frac{i}{128} I^c\left(N_c-3\right)\left(N_c-1\right)\left(N_c+5\right)\left(N_c+7\right)$ 	\\ \hline\hline
		\end{tabular}
	\end{table}

\begin{table}[h!]
	\begin{tabular}{c|l|l}\hline \hline
			$n_1$   $n_2$   $n_3$ & \;\;\;\;$G^{i a} \mathcal{P}_{n_3} G^{k d} \mathcal{P}_{n_2} G^{i a} \mathcal{P}_{n_1} G^{k d}$ & \;\;\;\;$\eps^{i j k} \eps^{a b c} G^{j b} \mathcal{P}_{n_3} G^{l d} \mathcal{P}_{n_2} G^{l d} \mathcal{P}_{n_1} G^{i a}$ \\ \hline
			$N$   $N$   $N$ & $\frac{\left(N_c+2\right)^4}{2304}$ & $-\frac{1}{144} I^c S^k\left(N_c+2\right)^4$ \\
			$N$  $N$   $\Delta$ & $\frac{1}{288}\left(N_c-1\right)\left(N_c+2\right)^2\left(N_c+5\right)$ & 0 \\
			$N$   $\Delta$  $N$ & $\frac{1}{288}\left(N_c-1\right)\left(N_c+2\right)^2\left(N_c+5\right)$ & $-\frac{1}{72} I^c S^k\left(N_c-1\right)\left(N_c+2\right)^2\left(N_c+5\right)$ \\
			$N$   $\Delta$  $\Delta$ & $\frac{5\left(N_c-1\right)\left(N_c+2\right)^2\left(N_c+5\right)}{1152}$ & 0 \\
			$\Delta$  $N$   $N$ & $\frac{1}{288}\left(N_c-1\right)\left(N_c+2\right)^2\left(N_c+5\right)$ & 0 \\
			$\Delta$   $N$   $\Delta$ & $\frac{\left(N_c-1\right)^2\left(N_c+5\right)^2}{2304}$ & $-\frac{1}{576} I^c S^k\left(N_c-1\right)^2\left(N_c+5\right)^2$ \\
			$\Delta$   $\Delta$   $N$ & $\frac{5\left(N_c-1\right)\left(N_c+2\right)^2\left(N_c+5\right)}{1152}$ & 0 \\
			$\Delta$  $\Delta$   $\Delta$ & $\frac{1}{288}\left(N_c-1\right)\left(N_c+2\right)^2\left(N_c+5\right)$ & $-\frac{1}{288} I^c S^k\left(N_c-1\right)\left(N_c+2\right)^2\left(N_c+5\right)$ \\
			$\Delta$   $\Delta_5$  $\Delta$ & $\frac{3}{256}\left(N_c-3\right)\left(N_c-1\right)\left(N_c+5\right)\left(N_c+7\right)$ & $-\frac{1}{192} I^c S^k\left(N_c-3\right)\left(N_c-1\right)\left(N_c+5\right)\left(N_c+7\right)$ \\
			\hline\hline
		\end{tabular}
	\end{table}

\vspace*{2cm}

\begin{table}[h!]
	\begin{tabular}{c|l|l}
		\hline\hline $n_1$ $ n_2 $ $ n_3$ & \;\;\;\;$\eps^{a b c} G^{i b} \mathcal{P}_{n_3} G^{k d} \mathcal{P}_{n_2} G^{i a} \mathcal{P}_{n_1} G^{k d}$ & \;\;\;\;$\eps^{i j k} \eps^{a b c} G^{j b} \mathcal{P}_{n_3} G^{l d} \mathcal{P}_{n_2} G^{i a} \mathcal{P}_{n_1} G^{l d}$ \\
		\hline $N$ $N$ $N$ & $-\frac{i}{1728} I^c\left(N_c+2\right)^4$ & $-\frac{1}{1296} I^c S^k\left(N_c+2\right)^4$ \\
		$N$ $N$ $ \Delta$ & $\frac{i}{432} I^c\left(N_c-1\right)\left(N_c+2\right)^2\left(N_c+5\right)$ & $-\frac{1}{648} I^c S^k\left(N_c-1\right)\left(N_c+2\right)^2\left(N_c+5\right)$ \\
		$N $ $\Delta$ $ N$ & $-\frac{i}{216} I^c\left(N_c-1\right)\left(N_c+2\right)^2\left(N_c+5\right)$ & $-\frac{1}{162} I^c S^k\left(N_c-1\right)\left(N_c+2\right)^2\left(N_c+5\right)$ \\
		$N$ $ \Delta $ $\Delta$ & $\frac{5 i}{1728} I^c\left(N_c-1\right)\left(N_c+2\right)^2\left(N_c+5\right)$ & $-\frac{5}{2592} I^c S^k\left(N_c-1\right)\left(N_c+2\right)^2\left(N_c+5\right)$ \\
		$\Delta$ $ N$ $ N$ & $-\frac{i}{216} I^c\left(N_c-1\right)\left(N_c+2\right)^2\left(N_c+5\right)$ & $-\frac{1}{162} I^c S^k\left(N_c-1\right)\left(N_c+2\right)^2\left(N_c+5\right)$ \\
		$\Delta$ $   N $ $ \Delta$ & $\frac{i}{3456} I^c\left(N_c-1\right)^2\left(N_c+5\right)^2$ & $-\frac{1}{5184} I^c S^k\left(N_c-1\right)^2\left(N_c+5\right)^2$ \\
		$\Delta $ $ \Delta$ $ N$ & $-\frac{5 i}{864} I^c\left(N_c-1\right)\left(N_c+2\right)^2\left(N_c+5\right)$ & $-\frac{5}{648} I^c S^k\left(N_c-1\right)\left(N_c+2\right)^2\left(N_c+5\right)$ \\
		$\Delta$ $ \Delta$ $ \Delta$ & $\frac{i}{432} I^c\left(N_c-1\right)\left(N_c+2\right)^2\left(N_c+5\right)$ & $-\frac{1}{648} I^c S^k\left(N_c-1\right)\left(N_c+2\right)^2\left(N_c+5\right)$ \\
		$\Delta $ $\Delta_5 $ $\Delta$ & $\frac{i}{128} I^c\left(N_c-3\right)\left(N_c-1\right)\left(N_c+5\right)\left(N_c+7\right)$ & $-\frac{1}{192} I^c S^k\left(N_c-3\right)\left(N_c-1\right)\left(N_c+5\right)\left(N_c+7\right)$ 
		\\ \hline\hline
	\end{tabular}
\end{table}

\begin{table}[h!]
\begin{tabular}{c|l|l}\hline \hline

		$n_1$ $ n_2 $ $n_3$ & \;\;\;\;$\eps^{i j k} \eps^{a b c} G^{l d} \mathcal{P}_{n_3} G^{j b} \mathcal{P}_{n_2} G^{i a} \mathcal{P}_{n_1} G^{l d}$ &\;\;\;\; $\eps^{a b c} G^{l d} \mathcal{P}_{n_3} G^{i b} \mathcal{P}_{n_2} G^{i a} \mathcal{P}_{n_1} G^{l d}$ \\ \hline
		$N $ $ N $ $ N$ & $-\frac{1}{1296} I^c S^k\left(N_c+2\right)^4$ & $\frac{i}{576} I^c\left(N_c+2\right)^4$ \\
		$N $ $N $ $\Delta$ & $-\frac{1}{648} I^c S^k\left(N_c-1\right)\left(N_c+2\right)^2\left(N_c+5\right)$ & 0 \\
		$N $ $\Delta $ $N$ & $-\frac{1}{2592} I^c S^k\left(N_c-1\right)\left(N_c+2\right)^2\left(N_c+5\right)$ & $-\frac{i}{576} I^c\left(N_c-1\right)\left(N_c+2\right)^2\left(N_c+5\right)$ \\
		$N $ $\Delta $ $\Delta$ & $-\frac{5}{648} I^c S^k\left(N_c-1\right)\left(N_c+2\right)^2\left(N_c+5\right)$ & 0 \\
		$\Delta $ $N $ $N$ & $-\frac{1}{648} I^c S^k\left(N_c-1\right)\left(N_c+2\right)^2\left(N_c+5\right)$ & 0 \\
		$\Delta $ $N $ $\Delta$ & $-\frac{25}{5184} I^c S^k\left(N_c-1\right)^2\left(N_c+5\right)^2$ & $-\frac{5 i}{1152} I^c\left(N_c-1\right)^2\left(N_c+5\right)^2$ \\
		$\Delta $ $\Delta $ $N$ & $-\frac{5}{648} I^c S^k\left(N_c-1\right)\left(N_c+2\right)^2\left(N_c+5\right)$ & 0 \\
		$\Delta $ $\Delta $ $\Delta$ & $-\frac{1}{648} I^c S^k\left(N_c-1\right)\left(N_c+2\right)^2\left(N_c+5\right)$ & $-\frac{i}{288} I^c\left(N_c-1\right)\left(N_c+2\right)^2\left(N_c+5\right)$ \\
		$\Delta$ $\Delta_5 $ $\Delta$ & $-\frac{1}{192} I^c S^k\left(N_c-3\right)\left(N_c-1\right)\left(N_c+5\right)\left(N_c+7\right)$ & $\frac{i}{128} I^c\left(N_c-3\right)\left(N_c-1\right)\left(N_c+5\right)\left(N_c+7\right)$ \\
		\hline\hline
	\end{tabular}
\end{table}
\end{center}

\end{widetext}

\newpage
$\quad$ \\

\bibliography{Refs-hlinked}
\end{document}